\documentclass[%
reprint,
amsmath,amssymb,
aps,
]{revtex4-2} 

\usepackage{color}
\usepackage{tikz}
\usetikzlibrary{shapes}

\usepackage{graphicx,subcaption,tabularx}
\usepackage{dcolumn}
\usepackage{bm}

\newcommand{\lb}{\left (}
\newcommand{\rb}{\right )}

\begin{document}
	
	
	\title{Undular Bores Generated by Fracture}
	
	\author{C.G. Hooper$^{1,2}$}
	\author{P.D. Ruiz$^1$}%
	\author{J.M. Huntley$^1$}%
	\author{K.R. Khusnutdinova$^2$}%

	\affiliation{%
		$^1$Wolfson School of Mechanical, Electrical and Manufacturing Engineering (WSMEME), Loughborough University, Loughborough LE11 3TU, UK}%
	\affiliation{%
		$^2$Department of Mathematical Sciences, Loughborough University, Loughborough LE11 3TU, UK}%
	
	\begin{abstract}
				\textit{Undular bores}, or \textit{dispersive shock waves}, are  non-stationary waves propagating as oscillatory transitions between two basic states, in which the oscillatory structure gradually expands and grows in amplitude with distance travelled. We demonstrate for the first time, using high-speed pointwise photoelasticity, the generation of undular bores in solid (polymethylmethacrylate) pre-strained bars by 
 natural and induced tensile fracture. 
 For the distances relevant to our experiments, the {\it viscoelastic extended Korteweg - de Vries (veKdV) equation} is shown to provide very good agreement with the key observed experimental features for suitable choice of material parameters, while some local features
 at the front of the bore are also captured reasonably well by the linearisation {\it near the nonzero pre-strain level}.
 The experimental and theoretical approaches presented open new avenues and analytical tools for the study and application of dispersive shock waves in solids.
	\end{abstract}
	
	\pacs{Valid PACS appear here}
	\maketitle
	
	

		\section{\label{sec:level2}Introduction}
	Dispersive shock waves (DSWs) have been observed and extensively studied in classical and quantum fluids, optics, and several other areas (e.g. \cite{
		GP, PS, WF, EH, 
		XCKMT, JSHW, AMS, 
		NFX} and references therein).
	Here we report an observation and modelling of  an undular bore generated by fracture in a pre-strained waveguide,
	which opens new opportunities for the studies of DSWs
	in solids, including possible adaptation to the processes at micro- and nano-scales.
	
	There exists experimental evidence that polymethylmethacrylate (PMMA) demonstrates weakly-nonlinear behaviour  for sufficiently strong deformations \cite{BBPSS, Sem_2017, WHG, Wang, DKSS}.
	We use PMMA because of its convenient optical properties, but the results are relevant to many elastic materials. 
	PMMA is the material of choice for windshields of most modern aircraft \cite{aircraft_1}, which must withstand the effect of the waves that occur as a result of possible fracture. 
Here 
	we are concerned only with the propagation of the waves, treating the measurement close to the fracture site as a given initial condition for the model equations.

Strain waves have been reported in PMMA and other materials, most commonly from Split Hopkinson Pressure Bar  (SHPB) and other impact tests \cite{Kolsky_1958, Acharya, Xue, Casem, Wang}. 
The 
strain waves produced in \cite{Kolsky_1958} show that an oscillatory transition region develops between the state of rest and the state of compressive strain due to geometrical dispersion in the waveguide. The qualitative features of an undular bore (expansion and growing amplitude of the oscillatory structure close to the transition) are clear in experiments in long metal waveguides 
\cite{Ren_army, SJ}. Similar waves in PMMA have been registered and studied in \cite{ZGK,H,Wang}.
The dominant engineering approaches have so far been based on Fourier analysis and numerical techniques, and this wave structure has not previously been identified as an undular bore.  Recognition that such waves are undular bores brings a significant benefit: the  
Korteweg - de Vries (KdV) type models (see \cite{OS, Samsonov, Porubov, Dai, GKS, GBK} and references therein) give access to additional analytical approaches.
While our modelling results will be useful in the context of these classical experiments, the main focus is on bores generated by 
fracture. To the best of our knowledge, such bores have not been experimentally observed or mathematically modelled despite the wide applicability of the problem formulation in natural and industrial settings,  as well as the relative ease of its laboratory implementation. 
The experiments extend the research in \cite{Miklowitz, Phillips, Kolsky_1973, KK_1976} by investigating the evolution of the release waves. 

	\section{Experiments}
	
		\subsection{Experiment 1: Natural Tensile Fracture}
	
 PMMA bars (3  $\times$ 10  $\times$ 750 mm$^3$) each cut from the same sheet of material were loaded at a constant strain rate of $3 \times 10^{-3}$ s$^{-1}$ until fracture, using a tensile testing machine (TTM, Instron 3345).  
	The bars were pre-notched with a knife blade run across both sides of the 10 mm wide sides of the sample  (so that the crack has to, ideally, propagate 3 mm to cause fracture), 100 mm away from one of the ends. 	
			The duration of the initial unloading is determined by the distance the crack tip propagates to cause fracture, and the speed at which it propagates.
		Crack propagation in viscoelastic materials has been extensively studied (e.g. \cite{Willis, MW, Fleck_2018}).
		Cracks in PMMA, once initiated, accelerate very rapidly to propagation speeds between
		$150$ m s$^{-1}$ and $300$ m s$^{-1}$ \cite{Takahashi_1984, Doll_1976, Sheng_1999, Wang_2019}. 
		The duration of the initial unloading due to the crack can therefore be estimated as being between $t = 10$ $\mu$s and $t = 20$ $\mu$s, but we note that it could take longer 
		should the crack not initiate uniformly across the full width of the notch tip.
	
\begin{figure}
		\begin{center}
			\includegraphics[width=8.1cm]{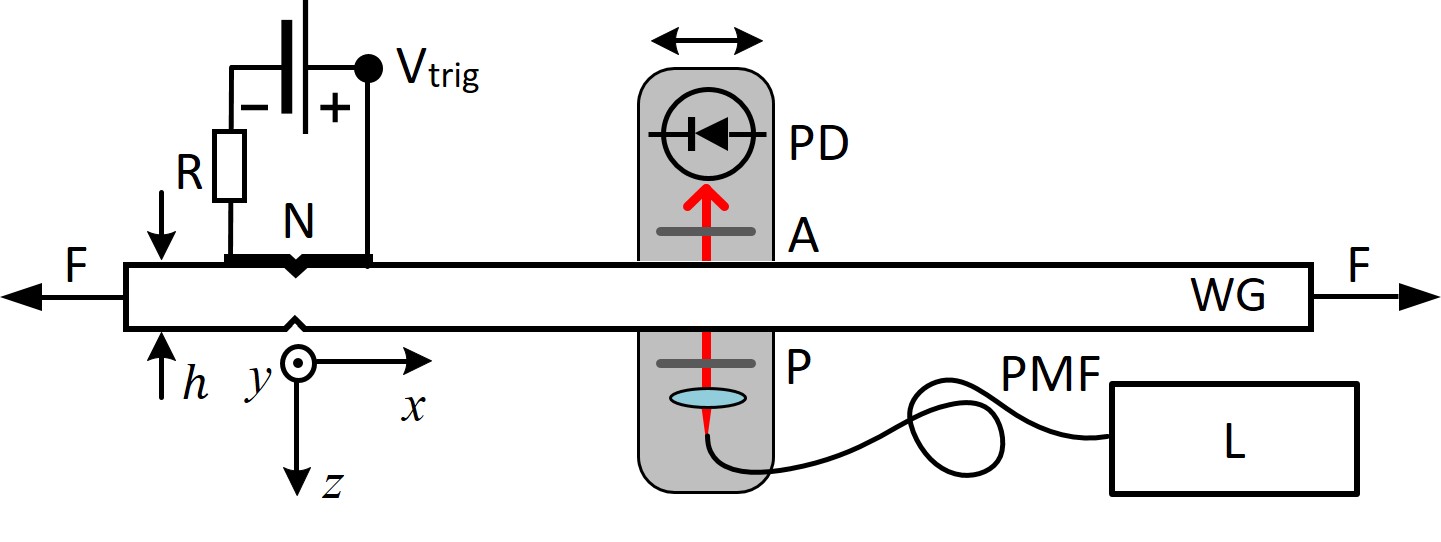}
		\end{center}
		\caption{Experimental setup, showing laser source (L), polarization maintaining fibre (PMF), polarizer (P), PMMA waveguide (WG) width $h = 3$ mm under tensile load (F) with a notch (N) at $x=0$, analyser (A), photodetector (PD). The triggering circuit recording $V_{trig}$ consists of a resistor (R), a 5V power supply and brittle conductive ink across the notch (thick).}
		\label{fig:exp}
	\end{figure}

	Once loaded into the TTM, the length of the sample between the grips was 700 mm and the notch was set 75 mm above the lower grip. 
	A schematic can be seen in Fig. \ref{fig:exp}, in a horizontal configuration. 
	
	As PMMA exhibits transient birefringence \cite{photoelasticity1}, a bright-field circular polariscope (CP) was used to measure the longitudinal strain in the bar. The CP consists of a He-Ne laser source (632.8 nm, 30 mW) coupled to a polarization maintaining single mode optical fibre (PMF, with FC-APC coupling), a short focal length lens that collimates the beam to a diameter of 1 mm, a circular polarizer (P), a circular analyser (A) and a photodetector (PD,  Thorlabs PDA36A2, 350 - 1100 nm, 12 MHz bandwidth). The optical setup is mounted on a platform (indicated in shaded grey in Fig. \ref{fig:exp}) that can slide along the PMMA bar.
	The use of a PMF ensures that environmental vibrations do not introduce strain-induced birefringence in the fibre, 
	and intensity changes measured by the photodetector are only caused by changes of strain in the bar. A simple triggering circuit was made by using brittle conductive ink across the notch and connecting it to a 1 k$\Omega$ resistor and a 5 V DC power supply. {\color{black}The circuit would open as soon as the crack reached one of the four corners of the two notches and the bar becomes discontinuous there.} The photodetector signal and the voltage $V_{trig}$ are measured with a digital oscilloscope (not shown). Data acquisition is triggered by a rising edge in $V_{trig}$ at fracture, which is taken as time $t=0$. 
	
 \subsection{Experiment 2: Induced Tensile Fracture} 
 PMMA bars (3 $\times$ 10 $\times$ 770 mm $^3$) cut from the same sheet as bars used in Experiment 1 were placed into the TTM and loaded at a constant strain rate of $1 \times 10^{-3}$ s$^{-1}$ until a uniaxial load of 1500 N was applied. The length of the sample between the grips was 720 mm. Fracture was then induced 1 second later by pressing a blade against the sample, 50 mm above the lower grip, and running it across the 10 mm width. The same arrangement as used in Experiment 1 was used to record the intensity (see Fig. \ref{fig:exp}). This 
 	method of fracture allowed us to achieve greater pre-strains and strain rates at fracture.

\subsection{Strain evaluation}
	The light intensity at the photodetector is given in our setting by
	\begin{equation}
	I = I_0\cos^2\left(\frac{\pi h (\sigma_x - \sigma_y)}{f_\sigma} \right ),
	\label{eqn:id}
	\end{equation}
	where $I_0$ is the intensity of the laser beam entering the sample, $h$ is the sample thickness, $\sigma_{x, y}$ are the respective values of stress and $f_\sigma = (2.02 - 2.30) \times 10^5$ Pa$\times$m/fringe is the fringe constant of the material  \cite{photoelasticity}.
Under uniaxial stress loading, $\sigma_y = 0$, and the  longitudinal strain $e_x$ and stress $\sigma_x$ are related by Hooke's law 
	$ \sigma_x = E e_x$, consistent with the assumptions that waves are long compared to $h$ and weakly nonlinear, made below. Here, $E$ is Young's modulus. 
	When equation (\ref{eqn:id}) is rearranged to obtain $e_x$, $N \pi$ should be added to the recovered phase, where $N$ is the (integer) fringe order.

\section{Results}
		\subsection{Natural Tensile Fracture}
		Over a series of 12 experiments, the CP was positioned between 0.05 m and $0.50 \text{ m}$ from the notch 
		with 0.05 m increments. 
		Details of the measurements for each test are given Table I.
\begin{table}[h]
	\begin{ruledtabular}
	\centering
	\begin{tabular}{cccc}
		\multicolumn{1}{l}{\textbf{Distance (m) \hspace{0.05cm} }} & \multicolumn{1}{l}{\textbf{Pre-strain \hspace{0.05cm} }}& \multicolumn{1}{l}{\textbf{Post-strain \hspace{0.05cm} }} & \multicolumn{1}{l}{\textbf{Load (kN)}} \\ \cline{1-4}
		0.05     \hspace{0.5cm}                                               & 0.0121       \hspace{0.5cm}             &0.0013                      & 0.99                                                \\
		0.05       \hspace{0.5cm}                                             & 0.0125        \hspace{0.5cm}                &0.0013                & 0.99                                                \\
		0.10      \hspace{0.5cm}                                              & 0.0099         \hspace{0.5cm}             &0.0013                    & 0.88                                               \\
		0.10      \hspace{0.5cm}                                             & 0.0111           \hspace{0.5cm}              &0.0015                & 0.91                                              \\
		0.15      \hspace{0.5cm}                                              & 0.0108          \hspace{0.5cm}               &0.0014                 & 0.92                                                \\
		0.20       \hspace{0.5cm}                                            & 0.0123           \hspace{0.5cm}               &0.0015               & 1.00                                                \\
		0.25       \hspace{0.5cm}                                            & 0.0096           \hspace{0.5cm}                &0.0013              & 0.86                                               \\
		0.30       \hspace{0.5cm}                                             & 0.0126          \hspace{0.5cm}                &0.0014              & 1.02                                                \\
		0.35		\hspace{0.5cm}				      	& 0.0100			\hspace{0.5cm}			&0.0013 		& 0.84						\\
		0.40        \hspace{0.5cm}                                            & 0.0090          \hspace{0.5cm}               &0.0010                 &0.79                                                \\
		0.45        \hspace{0.5cm}                                            & 0.0111           \hspace{0.5cm}               &0.0012               & 0.91                                                \\
		0.50        \hspace{0.5cm}                                           & 0.0103            \hspace{0.5cm}               &0.0015             & 0.82       
	\end{tabular}
	\caption{
		The pre-strain, post-strain (temporary residual strain due to viscous relaxation) and load at fracture of the bars in the natural tensile fracture experiment at each recorded distance. The mean and standard deviation of the pre-strain are $\kappa = 0.0109 \pm 0.0012 $.}                                       
	\label{table:experiment}
		\end{ruledtabular}
\end{table}

\begin{table}[h]
	\centering
		\begin{ruledtabular}
	\begin{tabular}{cccc}
		\multicolumn{1}{l}{\textbf{Distance (m) \hspace{0.05cm} }} & \multicolumn{1}{l}{\textbf{Pre-strain \hspace{0.05cm} }}& \multicolumn{1}{l}{\textbf{Post-strain \hspace{0.05cm} }} & \multicolumn{1}{l}{\textbf{Load (kN)}} \\ \cline{1-4}
		0.05     \hspace{0.5cm}                                               & 0.0217       \hspace{0.5cm}             &0.0048                      & 1.47                                               \\
		0.05       \hspace{0.5cm}                                             & 0.0233        \hspace{0.5cm}                &0.0048                 & 1.49                                               \\
		0.10      \hspace{0.5cm}                                              & 0.0225         \hspace{0.5cm}             &0.0049                    & 1.49                                              \\
		0.10      \hspace{0.5cm}                                             & 0.0225          \hspace{0.5cm}              &0.0049                & 1.48
		                                            \\  
				0.20      \hspace{0.5cm}                                             & 0.0208          \hspace{0.5cm}              &0.0052                & 1.47 
		\\         
				0.30      \hspace{0.5cm}                                             & 0.0217          \hspace{0.5cm}              &0.0050                & 1.47 
		\\                                           
	\end{tabular}
	\caption{
		The pre-strain, post-strain (temporary residual strain due to viscous relaxation) and load at fracture of the bars in the induced tensile fracture experiment at each recorded distance. The mean and standard deviation of the pre-strain is $\kappa = 0.022 \pm 0.0008 $.}                                       
	\label{Tab:blade}
		\end{ruledtabular}
\end{table}

	The strain profiles are shown in Fig. \ref{fig:exp_layer_comp} after they have been convolved with a 4 $\mu$s  time window (also performed with all subsequent experimental profiles presented)  and normalised against their pre-strain  $\kappa$ at which fracture occurred.  The profiles at 0.05 m and 0.10 m {\color{black}{(used later for parameter fitting)}} are each averages of two tests performed at the same distance.
	
	A period of nearly-constant strain $\kappa$ follows fracture whilst the release wave travels from the fracture site to the laser beam. No relaxation is observed during this period.  The strain then decreases from the peak strain $\sim$0.01 at a rate of $\sim$800 s$^{-1}$ over a time of $\sim$13 $\mu$s, which constitutes a 
	weakly-nonlinear  regime \cite{WHG, Wang, WMX}. 
	
	
	
	 The strain relaxes to a small positive value $\kappa_t$, as seen in the 0.05 m profile in Fig. \ref{fig:exp_layer_comp}. We refer to this temporary strain $\kappa _t$ as the post-strain. For the times relevant to our experiments, the post-strain is constant at each distance. It is calculated as the average strain from a $1\times 10^{-4}$ s window sufficiently far away from any oscillations. Total relaxation is eventually observed (in the order of seconds), thus there are no signs of plasticity.  
	Fluctuations in the strain behind the release wave with speeds of around 2200 m s$^{-1}$ and 1345 m s$^{-1}$ are observed as indicated in Fig. \ref{fig:exp_layer_comp}. 
	The slower wave has been confirmed as a shear wave by using high-speed digital image correlation and the grid method \cite{Grid_Method}. The peak strain rate during the high to low transition in the oscillatory part of the bore is $\sim$200 s$^{-1}$. 
	
	At each distance oscillations can be seen to emerge at the bottom of the release wave. 
	The shaded region in Fig. \ref{fig:exp_layer_comp} contains both the longitudinal and slower moving shear waves, also emitted during the fracture.

\subsection{Induced Tensile Fracture}

Over a series of 12 experiments, the CP was positioned between 0.05 m and 0.30 m from the fracture site. When a load of 1500 N was applied to the bars, the corresponding strain was $\sim$$0.022$, which constitutes a more nonlinear regime than encountered in the natural tensile fracture experiment (e.g. \cite{WHG, Wang, WMX}).  The strain rate of the release wave varied between tests due to the different loads being applied at the fracture site,  and a sample of 6 experiments with similar conditions at fracture was analysed.   The highest rates at each distance were $\sim$$2000$ s$^{-1}$.
The results are shown in Fig. \ref{fig:exp_blade}. The profiles at 0.05 m and 0.10 m used for parameter fitting are each averages of two tests performed at the same distance. Both pre-strain and strain rate here are around double those observed in the natural tensile fracture experiment. Other details are shown in Table \ref{Tab:blade}. 

Once again, longitudinal oscillations develop at the bottom of the release wave at each distance, and are clearly seen as the wave emerges out of the shaded region containing both the longitudinal and shear waves. The oscillations grow in amplitude and duration with propagation distance. 
It appears that in this regime of larger pre-strain and larger strain rate, the longitudinal oscillations emerge closer to the fracture site and with greater amplitude than with the lower rates produced by natural tensile fracture. The longitudinal oscillations behind the release wave can clearly be seen in the 0.20 m and 0.30 m profiles in Fig. \ref{fig:exp_blade} when there is sufficient distance between the release wave and the leading shear wave.

On comparison of tests with similar pre-strains and strain rates, 
it was observed that at distances relevant to both experiments the post-strain level typically gently increased with propagation distance. 

	\begin{figure}
	\centering
	\includegraphics[width=8cm]{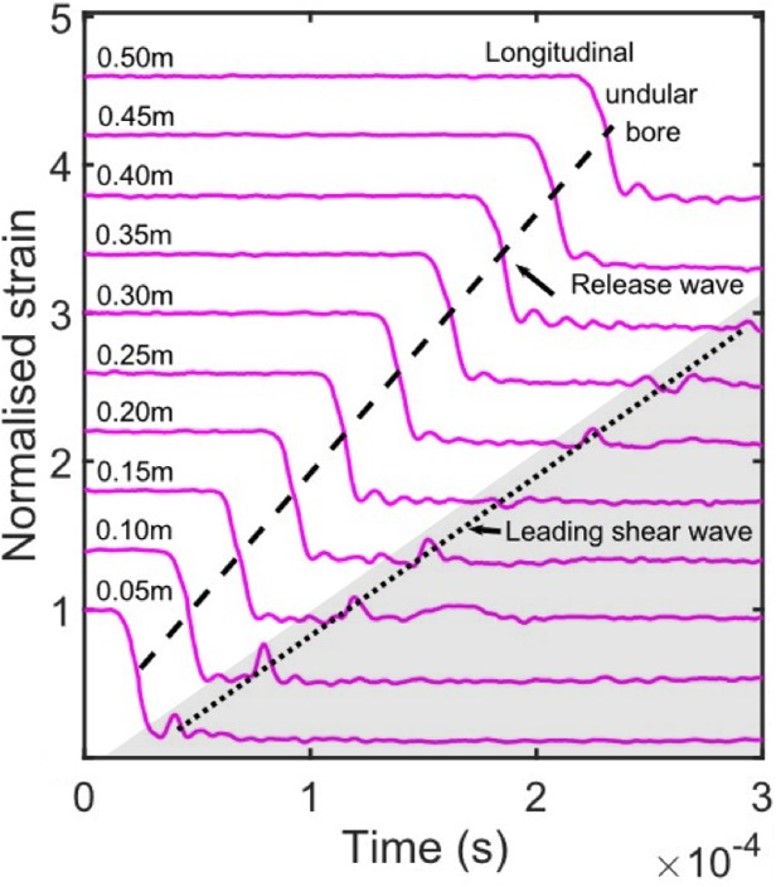}
	\caption{Experimental strain profiles at each recorded distance of the natural tensile fracture experiments. Profiles are normalised against their pre-strain. Vertical spacing of 0.4 is used to separate each profile.}
	\label{fig:exp_layer_comp}
\end{figure}

		\begin{figure}[h!]
			\centering
			\includegraphics[width=8.2cm]{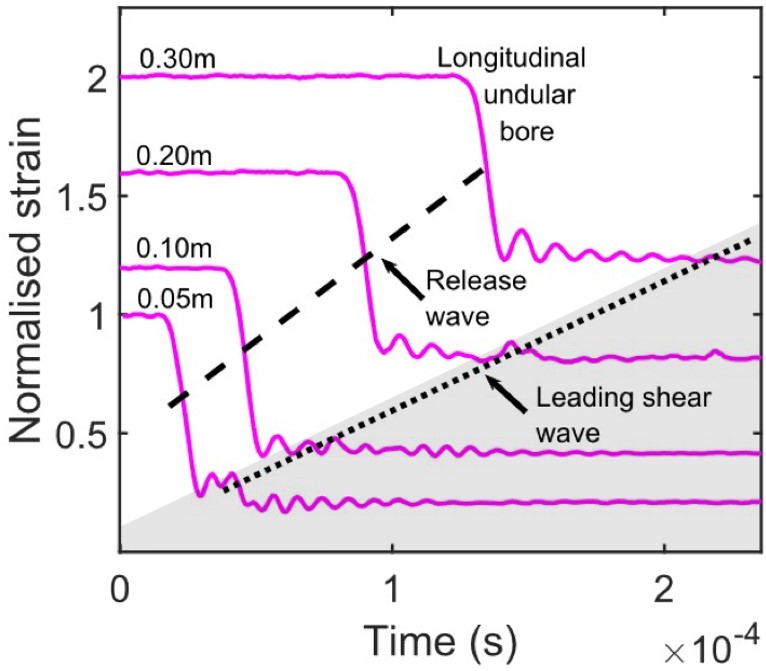}
			\caption{Experimental strain profiles at each recorded distance of the induced tensile fracture experiments. Profiles are normalised against their pre-strain. Vertical spacing of 0.4 is used to represent an increase of 0.10 m.}
			\label{fig:exp_blade}
		\end{figure}

The non-stationary structure developing in the faster moving longitudinal wave, as seen clearly in Fig. \ref{fig:exp_blade} (and also in Fig. \ref{fig:exp_layer_comp}) demonstrates  qualitative behaviour of an undular bore: it propagates as an oscillatory  transition between the levels of the pre- and post- strain, and the oscillatory structure gently expands and grows in amplitude with propagation distance. 


The features are similar to those in SHPB tests \cite{Kolsky_1958, Ren_army} discussed in the Introduction, but the bore develops here as a result of fracture. The quantitative characterisation  will be given and compared with the results of  linear and nonlinear modelling in the following sections.


	\section{Modelling}\label{sec:Modelling}
	
	 \subsection{Basic Elastic Modelling}
		The waveguide is an isotropic elastic
		bar of rectangular cross section $S = \{-b_1\leq y\leq
		b_1;-b_2\leq z\leq b_2\}$.  We assume that the bar is in the initial equilibrium state and introduce the Lagrangian Cartesian
		coordinates $(x, y, z)$, where $Ox$ is directed along the bar
		through the centre of each cross section area $S$.	We consider the action functional
		$$\Sigma=\int_{t_{0}}^{t_{1}} \int_{V }^{{}}L
		\ dV \  dt, 
		$$
		where $L = L({\bf U,U}_{t},{\bf U}_{x}\ldots x,t)$ is the Lagrangian density per unit volume, $t$ is time, $V$ is a space
		domain occupied by the waveguide, $\mathbf{U} =\{u,v,w\} $ is the displacement vector in the coordinates $(x, y, z)$. 

				The Lagrangian density $L$  is given by the difference of kinetic  $T$ and potential $\Pi $ energy densities, 
		$$
		L=T-\Pi =~\rho (\partial \mathbf{U}/\partial t)^{2}/2-\rho \Pi
		(I_{k}),
		$$
		where $\rho $ is the material density, and $I_{k}=I_{k}(\mathbf{C})$ are the
		invariants of Cauchy-Green's deformation tensor 
		\begin{eqnarray*}
			&&\mathbf{C}=[\mathbf{%
				\nabla U}+(\mathbf{\nabla U})^{T}+(\mathbf{\nabla U}%
			)^{T}\mathbf{\nabla U}]/2, \quad \mbox{where} \\
			&&I_{1}=tr\text{ }\mathbf{C},\text{ } \\
			&&I_{2}=(1/2)[(tr\text{ }\mathbf{C})^{2}-(tr%
			\text{ }\mathbf{C}^{2})],\text{ } \\
			&&I_{3}=\det \mathbf{C.}
		\end{eqnarray*}

	The quadratic and cubic nonlinearity were shown to become noticeable in PMMA at strains around $\sim$0.01 in \cite{WHG, Wang}.  Therefore, the derivation of  a Boussinesq-type equation for the long weakly-nonlinear longitudinal waves developed in \cite{KS} was extended to take into account the cubic nonlinearity. 
		
		The 9-constant Murnaghan model \cite{Murnaghan} 
		is used to describe the potential energy of compressible isotropic elastic materials,
		\begin{eqnarray*}
		\Pi &=& (\lambda +2\mu )I_{1}^{2}/2-2\mu
		I_{2}+(l+2m)I_{1}^{3}/3-2mI_{1}I_{2} \nonumber  \\
		&+& nI_{3} + \nu_1 I_1^4 + \nu_2 I_1^2 I_2 + \nu_3 I_1 I_3 + \nu_4 I_2^2,
		\label{mur}
		\end{eqnarray*}
		were, $\lambda$ and $\mu$ are Lame's coefficients, and $l, m, n, \nu_i$, where $i = \overline {1,4}$
		are Murnaghan's moduli.

			The model equation for the long weakly-nonlinear longitudinal waves is obtained in the approximation of
		the planar cross section hypothesis and approximate relations
		for the transverse displacements \cite{volterra56}:
		$$u = u(x,t), \text{ }v = -y \nu u_{x}, \text{ }w = -z \nu u_{x},$$ 
		where $\nu$ is Poisson's ratio.
		We assume the same scaling as in the asymptotic theory developed for waveguides with cylindrical geometry in \cite{GKS, GBK}. Then, up to appropriate quartic terms,
	\begin{eqnarray*}
		L &=&  \frac{\rho}{2}u_t^2 - \frac{1}{2}Eu_x^2 +  \frac{\rho  }{2}(y^2 + z^2) \nu^2 u_{xt}^2 \\
		&-& \frac{1}{2}\left[ \frac{\beta}{3}u_x^3 + \mu\nu^2(y^2+z^2)u_{xx}^2 \right] \\
		&-& \frac{1}{2} \left[ \frac{\gamma}{6}u_x^4 + \left(y^2 + z^2 \right) \gamma_1u_xu_{xx}^2 \right],
		\label{L}
	\end{eqnarray*}
	where the coefficients $\beta$, $\gamma$ and $\gamma_1$  are given by
	\begin{eqnarray*}
		\beta &=&
		3E+2l +4m -12l\nu +6(n-2m+4l)\nu^2 +\mathcal{O}(D\nu^3),\\
		\gamma &=& \frac{E}{8} + \frac{l}{2}+m + \nu_1 -2(l + 4\nu_1 + \nu_2)\nu \\
		&+& \left( \frac{E}{4} + 3l - m +\frac{n}{2} +24\nu_1 + 9\nu_2 + \nu_3 + 4\nu_4 \right) \nu^2 \\
		&+& \mathcal{O}(D\nu^3),\\[2ex]
		\gamma_1 &=& \left( \frac{E + m}{2}\right)\nu^2 +\mathcal{O}(D\nu^3), 
\end{eqnarray*}
		where $D = \text{max} \{E, l, m, ..., \nu_4 \}$. Thus, on taking terms up to ${\mathcal O} (D \nu)$, we have that $\gamma_1 = {\mathcal O} (D \nu^2)$, treating $\nu$ as a small parameter.
		
			The Euler-Lagrange equation $${\displaystyle \frac{\delta \mathcal{L}}{\delta u} = 0},$$ where
		$\displaystyle{\mathcal{L} = \int_{S} L dS}$ (Lagrangian density per unit length)  is regularised to remove the short-wave instability  \cite{KSZ} and differentiated with respect to $x$, yielding
 an extended Boussinesq-type equation for the leading order longitudinal strain $e = u_x$:

		\begin{equation}
		e_{tt} - c_0^2 e_{xx} = \frac{\beta}{2\rho}(e^2)_{xx} +   \frac{\gamma}{3\rho}(e^3)_{xx} + \delta^2 e_{ttxx},
		\label{eqn:DDE_strain} 
		\end{equation}
		where 
		$c_0=\sqrt{\frac{E}{\rho}}$ 
		and  $c_1= {\displaystyle \frac{c_0}{\sqrt{2(1+\nu)}}}$  are the linear longitudinal and shear wave velocities, and
		\begin{align*}
			 \delta^2 =  \frac{(b_1^2 + b_2^2) \nu^2}{3}  \left ( 1 - \frac{c_1^2}{c_0^2} \right ) 
			 = \frac{(b_1^2 + b_2^2) \nu^2 (1 + 2\nu)}{6(1+\nu)}.
			 \label{eqn:delta}
		\end{align*} 
		
		%
		In non-dimensional variables 
				\begin{equation}
			\tilde e = e/e_0, \hspace{0.3cm} \tilde t = t /T_0, \hspace{0.3cm} \tilde x = x/L_0,
			\label{ndv}
		\end{equation}
		where $e_0$ and $L_0$ are the characteristic amplitude and wave length, and $T_0 = L_0/c_0$, equation \eqref{eqn:DDE_strain} takes the form (omitting the tildes) 
		\begin{equation*}
		e_{tt} - e_{xx} = \epsilon \left[ \frac 12 \bar \beta  (e^2)_{xx}  + \bar{\gamma} (e^3)_{xx} + \bar{\delta}^2 e_{ttxx} \right], 
		\end{equation*}
		where 
		\begin{equation*}
		\epsilon = \frac{e_0 \lvert \beta \rvert}{E},\hspace{0.3cm} \bar \beta = {\rm sgn}\ \beta, \hspace{0.3cm}\bar{\gamma} = \frac{\gamma e_0}{3\lvert \beta \rvert}, \hspace{0.3cm}\bar{\delta}^2 = \frac{\delta^2 E}{L_0^2 e_0 \lvert \beta \rvert}.
		\end{equation*}
		We take $e_0$ to be the pre-strain level $\kappa$, and $L_0$ to be the characteristic length of the release wave.
			
			We look for a solution to this equation in the form of an asymptotic multiple-scale expansion
		\begin{equation*}
			e(x,t) = f(\xi, X) + \epsilon f^{(1)}(\xi,X) + \mathcal{O}(\epsilon^2),
			\label{ae}
		\end{equation*}
		where $\xi = x -  t$ 
		and $X = \epsilon x$. Thus, the waves propagate with the speed close to the linear longitudinal wave velocity $c_0 = \sqrt{\frac E \rho}$.
		
		The equation is satisfied at leading order, whilst at $O(\epsilon)$ we have
		\begin{equation*}
		f_X + \frac {\bar \beta}{2}  ff_\xi +  \frac{3\bar{\gamma}}{2}f^2f_\xi + \frac{ \bar{\delta}^2}{2}f_{\xi\xi\xi} =0.
		\label{eqn:G_nd}
		\end{equation*}
		Returning to dimensional variables we obtain 
		the Gardner equation  
		\begin{equation}
			e_x + \frac{1}{c_0}e_t - \frac{\beta}{2\rho c_0^3} ee_t  - \frac{\gamma}{2\rho c_0^3} e^2e_t -\frac{\delta^2}{2c_0^3}e_{ttt} = 0,
			\label{Gardner}
		\end{equation}
		where $\displaystyle{\frac{\beta}{2\rho c_0^3}}$ and $\displaystyle{ \frac{\gamma}{2\rho c_0^3}}$ are the quadratic and cubic nonlinearity coefficients, and $\displaystyle{\frac{\delta^2}{2c_0^3}}$ is the dispersion coefficient.
		
		It is worth noting that the strains encountered in our experiments are 2 orders of magnitude higher than the amplitudes of solitary waves in \cite{Sem_2017, BBPSS, DKSS, Samsonov_2008}. It is well known that small amplitude solitary waves of the Gardner equation are very close to the solitary waves of the Korteweg - de Vries equation, while extended models are better at describing waves of moderate amplitude  (see \cite{GBK} and references therein).	
		
		The linear approximation is obtained by linearising the Gardner equation near the nonzero pre-strain level $e = e_0 = const$, which can be formally viewed as changing the value of $E$ and letting
$
\beta = \gamma = 0.
$	
We note that the acoustoelastic effects (changes to the properties of the linear waves in pre-deformed media) have been extensively studied  (e.g. \cite{KH,DS,Pau,PNS}
and references therein). 	
The Korteweg - de Vries (KdV) model is obtained by considering the approximation
up to and including the quadratic terms.
We also note that the physically linear and geometrically nonlinear Saint-Venant-Kirchhoff (SVK) approximation is obtained by letting the Murnaghan moduli equal to zero, i.e.
$
l = m = n = 0,
$
which implies a particular case of the Gardner equation with 
\begin{equation}
\beta = 3 E, \quad \gamma = \frac{E}{8} (1 + 2 \nu^2).
\label{eqn:SVK_coeffs}
\end{equation}
The general Gardner equation (\ref{Gardner}) with independent parameters $\beta$ and $\gamma$ accounts for both the geometrical and physical sources of nonlinear corrections (up to cubic corrections). We also note that the dispersion coefficient does not depend on any nonlinear moduli.	 

	\subsection{Viscoelastic Corrections}
				Describing viscoelastic properties of polymers is a challenging task, and there are many different approaches.  Among the frequently used approaches we note the introduction of complex moduli and advanced computational algorithms accounting for attenuation and dispersion (e.g. \cite{ZG, BRY, BCGLM}), and models generalising Maxwell-Voigt-type elements (e.g. \cite{Wang_1993, W_2019, Harrigan_2012, Tamaogi_2013}). Other approaches to the modelling of viscoelasticity have been recently discussed,  in different contexts, in \cite{DPAP, DPNS, BDP} (see also the references therein).
			The possibility to adequately describe the dynamic behaviour of PMMA with the help of effective elastic moduli in intermediate regimes of  impact tests was reported in \cite{ZGK}. Waves very similar to undular bores generated by fracture in our study have been reported in impact tests with PMMA (although not recognised as undular bores), and the important role of the geometrical dispersion in situations when the wavelength is only moderately longer than the cross-sectional dimensions, as is the case in our study, has been established \cite{ZGK, H} (see also the references therein).
			The strain rates encountered in our experiments, and the initial wave profiles generated by tensile fracture \cite{Miklowitz, Phillips, Kolsky_1973, KK_1976}, are similar to those found in impact tests with PMMA. Therefore, it is natural that we also see the developing undular bores in our experiments (at room temperature, and as an intermediate regime). 
		 
				To include viscous effects in the model equation, here we use a spring and dashpot model consisting of a nonlinear spring of modulus $E_0$ and two Maxwell elements all in parallel with each other which was shown to be a good model for PMMA over a wide range of frequencies  (see \cite{Wang_1993, W_2019, Harrigan_2012} and references therein). In terms of our model, the effective constitutive equation is then
				\begin{align}
					\sigma &= E_0e + \frac{\beta}{2}e^2 + \frac{\gamma}{3}e^3 + \rho\delta^2e_{tt} \nonumber \\
					&+ E_1 \int_0^t \dot{e}(\tau) \textrm{exp}\left( \frac{t-\tau}{\theta_1}\right) d\tau \nonumber \\
					&+ E_2 \int_0^t \dot{e}(\tau) \textrm{exp}\left( \frac{t-\tau}{\theta_2}\right) d\tau,
					\label{eqn:ZWT_diss}
				\end{align}
				where we have included the dispersive term due to the geometrical dispersion. The first integral term describes the viscoelastic response of the material at low strain rates and has modulus and relaxation time $E_1$ and $\theta_1$ respectively. The second integral term describes the viscoelastic response of the material at high strain rates and has modulus and relaxation time $E_2$ and $\theta_2$ respectively.
				
				The slow relaxation time $\theta_1$ is of the order of seconds in accordance with the observed experimental time taken for total relaxation to occur. Thus $\theta_1$ is much larger than the times relevant to our experiments, and the first integral term reduces to a linear spring with modulus $E_1$. Then \eqref{eqn:ZWT_diss} becomes
				\begin{align}
					\sigma &= E_ae + \frac{\beta}{2}e^2 + \frac{\gamma}{3}e^3 + \rho\delta^2e_{tt} \nonumber \\
					&+ E_2 \int_0^t \dot{e}(\tau) \textrm{exp}\left( \frac{t-\tau}{\theta_2}\right) d\tau,
					\label{eqn:ZWT_diss_2}
				\end{align} 
				where $E_a = E_0 + E_1$. 
				On substituting \eqref{eqn:ZWT_diss_2} into the equation of motion 
				\begin{equation*}
					\rho e_{tt} = \dfrac{\partial^2 \sigma}{\partial x^2},
				\end{equation*}
				we have
				\begin{align}
					\rho e_{ttt} &+ \frac{\rho}{\theta_2}e_{tt} = E_A e_{xxt} + \left( \frac{\beta}{2}e^2 + \frac{\gamma}{3}e^3 + \rho\delta^2 e_{tt} \right)_{xxt} \nonumber \\
					&+\frac{1}{\theta_2}\left( E_a e + \frac{\beta}{2} e^2 + \frac{\gamma}{3}e^3 + \rho\delta^2e_{tt}\right)_{xx},
					\label{eqn:ZWT_1}
				\end{align}
				where $E_A = E_0 + E_1 + E_2$.  We now non-dimensionalise \eqref{eqn:ZWT_1} in the same spirit as with \eqref{eqn:DDE_strain} by using the nondimensional variables \eqref{ndv} with $T_0 = L_0/c$ where $c^2 = E_A/\rho$.
				This gives, on omitting tildes,
				\begin{align}
					\left(e_{tt} - e_{xx}\right)_t &= \hat{\epsilon} \left( -e^2 + \hat{\gamma}^2e^3 + \hat{\delta}^2e_{tt}\right)_{xxt} \nonumber \\
					&- \tilde{\epsilon} \left( e_{tt} - \frac{c_a^2}{c^2}e_{xx} \right) \nonumber \\
					&+ \hat{\epsilon} \tilde{\epsilon} \left( -e^2 + \hat{\gamma}^2e^3 + \hat{\delta}^2e_{tt} \right)_{xx},
					\label{eqn:ZWT_2}
				\end{align}
				where $c_a^2 = E_a/\rho$. We have introduced two small parameters  
				\begin{equation*}
					\hat{\epsilon} = \frac{e_0 \lvert \beta \rvert}{2\rho c^2}, \hspace{0.5cm} \tilde{\epsilon} = \frac{T_0}{\theta_2},
				\end{equation*}
				and constants 
				\begin{equation*}
					\hat{\gamma}^2 = \frac{2\gamma e_0}{3 \lvert \beta \rvert}, \hspace{0.5cm} \hat{\delta}^2 = \frac{2\delta^2 E_A}{e_0  \lvert \beta \rvert L_0^2 }.
				\end{equation*}
				We note that on using the values of $E_G$, $\beta_G$ and $\gamma_G$ determined in the next section, and the value of $\theta_2 \sim 1 \times 10^{-4}$ s from \cite{Wang_1993} we have
				\begin{equation*}
					\hat{\epsilon} \sim 0.07, \hspace{0.1cm}\tilde{\epsilon} \sim 0.1,\hspace{0.1cm} \hat{\gamma}^2 \sim 0.6, \hspace{0.1cm}\hat{\delta}^2 \sim 0.01.
				\end{equation*}
				Keeping the small dispersive correction is important in order to describe the dispersive resolution of the gradient catastrophe (e.g. \cite{GP,PS,EH,AMS} and references therein).
				
				Looking for a solution to \eqref{eqn:ZWT_2} in the form 
				\begin{equation*}
					e(x,t) = f(\xi, X) + \hat{\epsilon} f^{(1)}(\xi,X) + \mathcal{O}(\hat{\epsilon}^2),
					\label{ae2}
				\end{equation*}
				where $\xi = x - t$ and $X = \hat{\epsilon}x$,
				at order $\hat{\epsilon}$ we have
				\begin{align*}
				&f_X - ff_{\xi} + \frac{3\hat{\gamma}^2}{2}f^2f_{\xi} + \frac{\hat{\delta^2}}{2}f_{{\xi}{\xi}{\xi}} + \frac{1}{2}\frac{\tilde{\epsilon}}{\hat{\epsilon}} \left(1 - \frac{c_a^2}{c^2}\right)f \nonumber \\
				&+ A(X)\xi + B(X) = 0.
				\label{eqn:visco_nd}
			\end{align*}
				We have neglected terms $\mathcal{O}(\tilde{\epsilon})$ but not $ \displaystyle{\mathcal{O}(\tilde{\epsilon}/\hat{\epsilon}) \sim \mathcal{O}(1)}$. The functions $A(X)$ and $B(X)$ result from integration with respect to $\xi$ twice. 
				From our experimental observations, and at relevant distances, the post-strain is a function of distance $x$ only, therefore $A(X) = 0$ to remove the dependence on time through the variable $\xi$.  At all distances relevant to our experiments we observe no relaxation of the pre-strain at times before the release wave passes through.
				Hence in the pre-strain region $(f = 1)$ we require that $\displaystyle B(X) =- \frac{1}{2}\frac{\tilde{\epsilon}}{\hat{\epsilon}} \left(1 - \frac{c_a^2}{c^2}\right)$ so that the pre-strain level at all distances from fracture remains constant until the release wave arrives. Hence we have
				\begin{equation}
					f_X - ff_{\xi} + \frac{3\hat{\gamma}^2}{2}f^2f_{\xi} + \frac{\hat{\delta^2}}{2}f_{{\xi}{\xi}{\xi}} + \frac{1}{2}\frac{\tilde{\epsilon}}{\hat{\epsilon}} \left(1 - \frac{c_a^2}{c^2}\right)(f-1) = 0.
					\label{eqn:visco_nd2}
				\end{equation}
				Returning to the original dimensional variables, \eqref{eqn:visco_nd2} becomes
				\begin{align}
					e_x &+ \frac{1}{c}e_t - \frac{\beta}{2\rho c^3} ee_t  - \frac{\gamma}{2\rho c^3} e^2e_t -\frac{\delta^2 E_A}{2\rho c^5}e_{ttt} \nonumber \\
					& + \frac{\mu}{2 c}(e-e_0) = 0,
					\label{eqn:visco}
				\end{align}
				where $$\displaystyle \mu = \left(1 - \frac{c_a^2}{c^2}\right)/\theta_2.$$
				Thus, we conclude that the leading-order viscoelastic correction, within the scope of the present model, is similar to the Rayleigh dissipation term of the models describing undular bores in fluids (e.g. \cite{CS}). Note that on setting $\displaystyle{ E_1 = E_2 = 0}$, equation \eqref{eqn:visco} reduces to the Gardner equation \eqref{Gardner}. 
\subsection{Higher-order Dispersive and Nonlinear Corrections}		
Finally, following from the recent systematic asymptotic derivation of the equation of motion for a longitudinal strain wave in a nonlinear elastic rod (circular cross section) \cite{GBK}, we introduce the next dispersive correction and two nonlinear terms  present at the same order as the cubic nonlinear term with coefficient $\gamma$ (see (\ref{L})), thus the full equation is given by the following {\it viscoelastic extended Korteweg - de Vries (veKdV) equation}
	\begin{align}
		e_x &+ \frac{1}{c_0}e_t - \frac{\beta}{2\rho c_0^3} ee_t  - \frac{\gamma}{2\rho c_0^3} e^2e_t -\frac{\delta^2}{2c_0^3}e_{ttt}  \nonumber \\
		& + \frac{\mu}{2 c_0}(e-e_0)  \nonumber \\
		& + a_1 e_{ttttt} + a_2e_te_{tt} + a_3 ee_{ttt}= 0,
		\label{5th_V_G}
	\end{align}
	for some parameters $a_1$, $a_2$ and $a_3$, which have been calculated for the rod in \cite{GBK} but are currently unknown for a bar of rectangular cross section.  We have also identified the effective modulus of the system with the dynamic Young's modulus $E_A = E$. 
	
	We will refer to equation \eqref{5th_V_G} with $\mu = 0$ as the extended Korteweg - de Vries (eKdV) equation. The eKdV equation has previously emerged and was extensively studied in the context of fluids (e.g. \cite{
	Slyunyaev99,
	Grimshaw02,
	Karczewska14,
	KST} and references therein).

\section{Parameter fitting}
We use results from both experiments to obtain suitable parameters suitable for each regime.
		Parameters that are the same in both regimes are the geometric parameters measured as $b_1 = 0.005$ m and $b_2 = 0.0015$ m, the density of the samples measured as $\rho = 1060$ kg m$^{-3}$ and $\nu = 0.34$, in agreement with the observed shear and longitudinal wave speeds.	
		
		The third-order Murnaghan's moduli of similar polymers have been measured at low strain rates in \cite{KH, S2020}. 
		Experiments with strain solitons 
		in \cite{BBPSS, Sem_2017} and 
		ultrasonic waves in
		\cite{polymer_testing} have given evidence that $\beta < 0$ for PMMA.
		Young's modulus of PMMA 
		increases with increasing strain rates \cite{WMX, Fabrice}. 
		In what follows we are not concerned with the expressions for the nonlinearity coefficients in terms of Murnaghan's moduli (since the latter are unknown for the conditions of our experiment), but instead use $\beta$ and $\gamma$ which matter the most at distances close to the fracture site, as the main effective fitting parameters, alongside $E$ (known from \cite{Fabrice}, and used as a control parameter).  We then fine-tune the solution by fitting the remaining parameters in the areas of their importance. 

		We first consider the results of the tensile fracture experiment (see Fig. \ref{fig:exp_layer_comp}) to provide parameters suitable for pre-strains and strain rates of $\sim$$1\%$ and $\sim$$800$ s$^{-1}$ respectively.

		

	%
	%

	The initial profile 
	 for all models 
	is approximated by a smooth decreasing step between the levels of $\kappa$ and $\kappa_t$  
	\begin{equation}
	e(x_0,t) = \kappa_t + \frac{\kappa-\kappa_t}{2}\left [ 1 - \text{erf} \lb \frac{t-\eta_1}{2 \tilde{L}} \rb \right ],
	\label{eqn:IC}
	\end{equation}
	where 
	$x_0 = 0.05$ m.  	We take $\kappa$ as the average of pre-strain from the entire set of experiments which is $\kappa = 0.0109$. The value of $\kappa_t$ is taken as the average of the post-strain in the 0.05 m curves only, as we observe that the level of post-strain changes with distance, thus $\kappa_t =  0.0013$.

The `slope' $\tilde L$ (transition time) and shift $\eta_1$ were found
	 by numerically fitting \eqref{eqn:IC} to the experimental strain at 0.05 m in the region $e \in [0.30,0.45]$ using the \emph{lsqnonlin} function in MATLAB \cite{MATLAB}. This region was chosen as the bore develops from it, and gave $\eta_1 = 2.3\times10^{-5}$ s and $\tilde L = 2.63\times10^{-6}$  s. 

A least squares fitting that used all experimental data was attempted, but proved unsuccessful as the results produced parameters giving no oscillations at relevant times. Therefore, the parameters $E$, $\beta$ and $\gamma$ 
	were determined by the following theoretical method.

\begin{figure}
	\centering
	\includegraphics[width=0.45\textwidth]{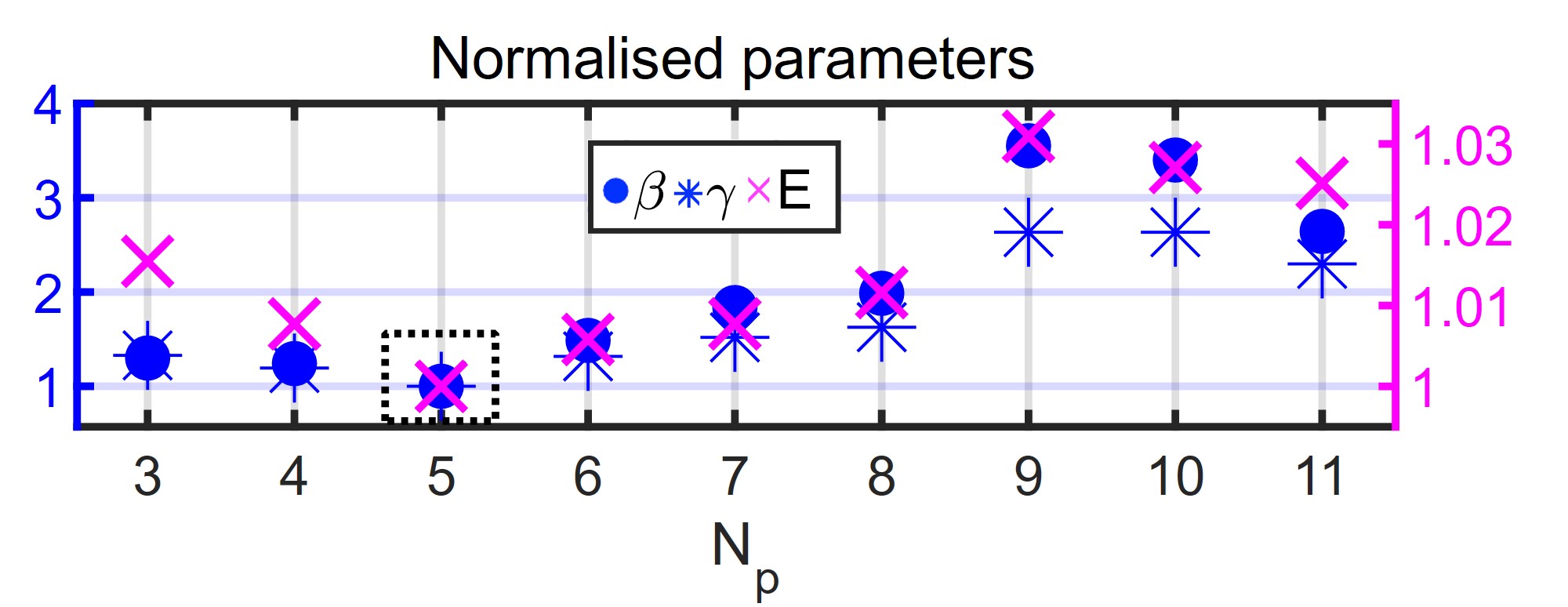}
	\caption{ The values of $E$ (right axis), $\beta$ and $\gamma$ (left axis)  on splitting the region $[0.30, 0.45]$ by $N_p$ points from natural tensile fracture. Values are normalised against  $E_G = 5.17$ GPa, $\beta_G = -74.4$ GPa and $\gamma_G = 5940$ GPa.}
	\label{fig:fit_tensile}
\end{figure}

	We assume that for the initial profile \eqref{eqn:IC} the initial evolution is governed by the hyperbolic equation 
	\begin{equation}
	e_x + \frac{1}{c_0} e_t - \frac{\beta}{2\rho c_0^3} ee_t  - \frac{\gamma}{2\rho c_0^3} e^2e_t = 0.
	\label{eqn:hyperbolic} 
	\end{equation}
	All other terms are assumed to be small close to the fracture site. Equation \eqref{eqn:hyperbolic} can be solved 
	by the method of characteristics. In particular we find that the wave speed at $e = const$  is given by
	\begin{equation*}
	\displaystyle \frac{dx}{dt} = \frac{2\rho c_0^3}{2\rho c_0^2 - \beta e - \gamma e^2}.
	\label{eqn:charac_speed}
	\end{equation*}
	By choosing three distinct points on the initial profile that match the experimental curve, we find the unknown parameters by requiring that 
	those points are mapped correctly onto the experimental strain profile at the next recorded distance (0.10 m). 
	We chose the initially fitted region $ \displaystyle{e \in [0.30, 0.45]}$ and divided this region into $N_p-1$ equal intervals. The end points of all intervals form a set of $N_p$ points. For each $N_p = \overline {3, 11}$, we found parameters for the $\displaystyle{C^{N_p}_k = \frac{N_p!}{k! (N_p-k)!}}$ possible triples of points and took the average.  The results are shown in Table \ref{tab:2} and also in	Fig. \ref{fig:fit_tensile}.

\begin{table}[h]
	\centering
		\begin{ruledtabular}
	\begin{tabular}{cccc}			
		\multicolumn{1}{l}{\textbf{$N_p$ \hspace{0.5cm}  }} & \multicolumn{1}{l}{\textbf{$E$ (GPa)\hspace{0.5cm} }} & \multicolumn{1}{l}{\textbf{$\beta$ (GPa)\hspace{0.5cm} }} & \multicolumn{1}{l}{\textbf{$\gamma$ (GPa)}} \\ \cline{1-4}
		3     \hspace{0.5cm}                                              & 5.25 \hspace{0.5cm}                                         &-98.8      \hspace{0.5cm}                                     &7720  \hspace{0.5cm}        \\    
		4        \hspace{0.5cm}                                        &   5.21 \hspace{0.5cm}                                      &-88.8      \hspace{0.5cm}                                     & 7370  \hspace{0.5cm}        \\   
		5       \hspace{0.5cm}                                                & 5.17   \hspace{0.5cm}                                       & -74.4   \hspace{0.5cm}                                      &5940  \hspace{0.5cm}        \\
		6   \hspace{0.5cm}                                              & 5.20    \hspace{0.5cm}                                     & -98.0     \hspace{0.5cm}                                     &8830   \hspace{0.5cm}       \\           
		7   \hspace{0.5cm}                                              & 5.21     \hspace{0.5cm}                                     & -113      \hspace{0.5cm}                                    &10900   \hspace{0.5cm}        \\  
		8   \hspace{0.5cm}                                               & 5.23     \hspace{0.5cm}                                     & -121        \hspace{0.5cm}                                  &11800   \hspace{0.5cm}        \\    
		9   \hspace{0.5cm}                                               & 5.33     \hspace{0.5cm}                                     & -196        \hspace{0.5cm}                                  &21100   \hspace{0.5cm}        \\   
		10   \hspace{0.5cm}                                               & 5.31     \hspace{0.5cm}                                     & -196        \hspace{0.5cm}                                  &20200   \hspace{0.5cm}        \\        
		11   \hspace{0.5cm}                                               & 5.30     \hspace{0.5cm}                                     & -171        \hspace{0.5cm}                                  &15700   \hspace{0.5cm}        \\ 		                             
	\end{tabular}
	\caption{The averaged values of $E$, $\beta$ and $\gamma$ found from fitting with different numbers of points $N_p$ using averaged 0.05 m and 0.10 m profiles from natural tensile fracture.}                                       
	\label{tab:2}
		\end{ruledtabular}
\end{table}

\begin{table}[t]
	\centering
		\begin{ruledtabular}
	\begin{tabular}{ccc}			
		\multicolumn{1}{l}{\textbf{$N_p$ \hspace{0.5cm}  }} & \multicolumn{1}{l}{\textbf{$E$ (GPa)\hspace{0.5cm} }} & \multicolumn{1}{l}{\textbf{$\beta$ (GPa) }}  \\ \cline{1-3}
		3     \hspace{0.5cm} & 5.13 \hspace{0.5cm}                                        &-36.2      \hspace{0.5cm}                                               \\    
		4        \hspace{0.5cm}                                        &   5.13 \hspace{0.5cm}                                      &-35.1     \hspace{0.5cm}                                             \\   
		5       \hspace{0.5cm}    & 5.13  \hspace{0.5cm}                                       & -34.2 \hspace{0.5cm}                                              \\
		6   \hspace{0.5cm}                                              & 5.13   \hspace{0.5cm}                                     & -34.2     \hspace{0.5cm}                                             \\           
		7   \hspace{0.5cm}                                              & 5.12     \hspace{0.5cm}                                     & -33.4      \hspace{0.5cm}                                          \\  
		8   \hspace{0.5cm}                                               & 5.12     \hspace{0.5cm}                                     & -33.4        \hspace{0.5cm}                                         \\    
		9   \hspace{0.5cm}                                               & 5.12    \hspace{0.5cm}                                     & -34.2       \hspace{0.5cm}                                     \\   
		10   \hspace{0.5cm}                                               & 5.13    \hspace{0.5cm}                                     & -34.9        \hspace{0.5cm}                                          \\        
		11   \hspace{0.5cm}                                               & 5.14    \hspace{0.5cm}                                     & -37.3       \hspace{0.5cm}                                          \\ 		                             
	\end{tabular}
	\caption{The averaged values of $E$ and $\beta$ (with $\gamma = 0$) found from fitting with different numbers of points $N_p$ using averaged 0.05 m and 0.10 m profiles from natural tensile fracture.}                                       
	\label{tab:kdv}
		\end{ruledtabular}
\end{table}

	All parameters have a 
	minimum at $N_p = 5$ shown by the dashed box in Fig. \ref{fig:fit_tensile}, indicating that the fitting errors are close to zero (i.e. in the vicinity of these values $dE \approx 0, d \beta \approx 0, d \gamma \approx 0$). The values of the parameters at $N_p = 5$ are $E_G = 5.17$ GPa, $\beta_G = -74.4$ GPa and $\gamma_G = 5940$ GPa. 

\begin{figure}
	\centering
	\includegraphics[width=9.1cm]{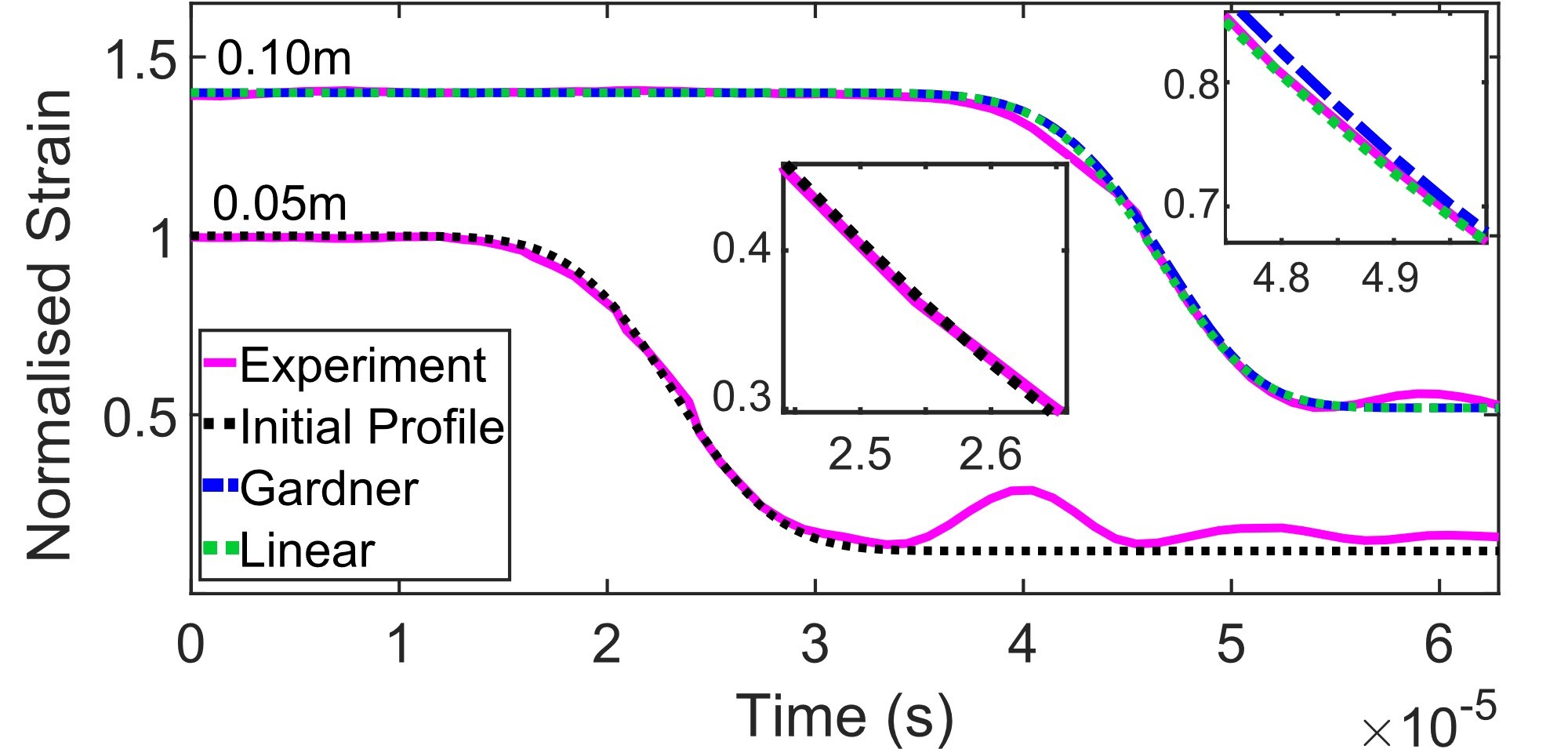}
	\caption{ The initial profile (0.05 m) and the Gardner \eqref{Gardner} and linearised Gardner \eqref{lin_G} solutions at 0.10 m with corresponding natural tensile fracture experimental
			profiles used to fit the initial condition and parameters
			of the models in the regime of $1\%$ strain and 800 s$^{-1}$ strain rate.}
	\label{fig:fit}
\end{figure}


			We must emphasise that these effective parameters are determined only by the first two strain profiles at $x = 0.05$ m and $0.10$ m, and there is no more fitting at subsequent distances up to $x = 0.50$ m.  
		Note that $\beta$ is negative and 
		all values of the control parameter $E$ found (to 1 decimal place) fall within the range of values $5.0-5.3$ GPa
		measured experimentally at similar strain rates to what we encounter in \cite{Fabrice}. 
		
		This was repeated with $\gamma = 0$ to give the analogue fitting method for the KdV model. Parameters were found for the $\displaystyle{C_k^{N_p} = \frac{N_p !}{k!(N_p-k)!}}$ possible pair of points and then averaged to give the values shown in Table \ref{tab:kdv}. A minimum is observed for both parameters around the points $N_p = 7, 8$,
			thus $E_{KdV} = 5.12$ GPa and $\beta_{KdV} = -33.4$ GPa. 
			On comparison of the KdV model solution to experimental results with these parameters we found the slope of the release wave to be too steep, and the cubic nonlinearity to be important. This is understandable since the  strains are much greater than those described by the KdV approximation. We  also note that our values are not too far from the  values of  $E$ and $\beta$ found from completely different compression strain wave experiments on PMMA: $E_{KdV} = 5.27$ GPa and $\beta_{KdV} = -15.9$ GPa (e.g. \cite{Sem_2017}). 
				
		For the linear and SVK models obtained from \eqref{Gardner}, the only unknown parameter is $E$. We fitted this numerically such that the solution at 0.10 m best matched the experimental data in the initially fitted region which gave $E_{lin} = 5.05$ GPa and $E_{SVK} = 4.96$ GPa. The resulting solution for these two models was similar.  
		
			We note here that the reduced value of $E_{lin}$ compared to $E_G$ is consistent with the linearisation of the Gardner equation {\it around the average value of the pre-strain} of the profiles at the distance $0.10$ m used to fit parameters of both models. Indeed, the average pre-strain of the 0.10 m profiles is $\bar{\kappa} = 0.0105$.
			Then, substituting $e = \bar{\kappa} + \bar e(x,t)$ into \eqref{Gardner} and linearising with respect to $\bar e$ gives 
			\begin{equation}
				\bar e_{x} + \left( \frac{1}{c_0} - \frac{\beta \bar{\kappa^2}}{2\rho c_0^3}   - \frac{\gamma \bar{\kappa}^2}{2\rho c_0^3}\right)\bar e_{t} - \frac{\delta^2}{2_0^3}\bar e_{{ttt}} = 0.
				\label{lin_G}
			\end{equation}
			The wave speed of this linearised Gardner equation is given by the reciprocal of the $\bar e_{t}$ coefficient as
				$$\displaystyle c_{lin} = \frac{2\rho c_0^3}{2\rho c_0^2 - \beta \bar{\kappa} - \gamma \bar{\kappa}^2}.$$
			On substitution of $E = E_G$, $\beta = \beta_G$ and $\gamma = \gamma_G$, we obtain
				$c_{lin} = \sqrt{ \frac{E}{\rho}}  = 2.18 \times 10^3, $
			which gives $E = E_{lin} = 5.05$ GPa, which is the same as was found above in the independent fitting. 

		We note the value of $E_{SVK}$ to be smaller than other values reported for PMMA in this regime and  $\beta$ to be positive from \eqref{eqn:SVK_coeffs}, whereas in our fitting $\beta$ was always negative. Thus, we rule out the KdV and SVK model from the forthcoming discussion, and concentrate on the extended KdV model as the one giving the best fit, and the linearised model (linearised near the nonzero pre-strain level) as a simple model which, at the distances relevant to our experiment, partially captures some features of the full model at the front of the bore. The initial profile and the solutions of the Gardner model \eqref{Gardner} and linearised near pre-strain Gardner model  (\ref{lin_G})  
		at 0.10 m are shown in Fig \ref{fig:fit}.
			
	On comparing the solution of the models obtained from the Gardner equation \eqref{Gardner} with parameters $E_G$, $\beta_G$ and $\gamma_G$ to corresponding results from our experiments (discussed later), we observed at all distances that the duration of the leading oscillation in the models, defined as the time between the first and second minima, was a bit smaller than that in experiments. Therefore we introduced the fifth derivative dispersive correction and fitted $a_1$ such that the duration of the model matched the experiment at the last recorded distance (0.50 m) which gave $a_1 =   1\times 10^{-28}$ s$^5$ m$^{-1}$. Finally $a_2$ was fitted such that the amplitude of the leading oscillation, defined as the vertical distance from the first minimum to the first maximum, matched the experimental amplitude, giving $a_2 = 5.7\times 10^{-14}$ s$^3$ m$^{-1}$. The parameter $a_3$ was found to offer no significant improvement to the solution in comparison to the experiments, so is omitted from the modelling, for simplicity. While there are no analytical expressions for these coefficients for the bar, the order of magnitude of the fitted coefficients is close to that for the known coefficients computed for a rod with the radius equal to the half the width of our bar.
	

	In nondimensional variables \eqref{ndv}, the fifth dispersion and $e_te_{tt}$ coefficients are
\begin{equation*}
	\bar{a_1} = -\frac{a_1 c_0^5 E}{L_0^4 e_0 \lvert \beta \rvert}, \quad \bar{a_2} = -\frac{a_2 c_0^3E}{L_0^2 \lvert \beta \rvert},
\end{equation*} 
respectively. Our assumptions can now be validated as we find that $\epsilon = 0.157$, $\bar \beta = -1$,  $\bar{\gamma} = 0.290$, $\bar{\delta}^2 = 0.005$, $\bar{a_1} = -0.0001$ and $\bar{a_2} = -0.047$ with $E_G$, $\beta_G$ and $\gamma_G$ ($e_0 \sim 0.0109$, $L_0 \sim 0.03$ m). We do not fit $\mu$ 
here as the spread of the experimental data was too large.	

We now turn attention to the induced fracture experiment. The smaller range of pre-strain and strain rate (i.e. better control of the conditions at fracture) allow us to fit the viscoelastic parameter $\mu$ by looking in the region of post-strain. Here, sufficiently far away from the release wave, oscillations and shear waves, all time derivatives are close to zero. Hence equation \eqref{5th_V_G} reduces to
\begin{equation}
	e_x + \frac{\mu}{2c}(e-e_0) = 0.
	\label{eqn:PostStrain}
\end{equation}
Solving \eqref{eqn:PostStrain} gives the strain in the post-strain region as
\begin{equation*}
	e = e_0 - A\textrm{exp}\left(-\frac{\mu}{2c}x \right),
\end{equation*}
where $A$ is a constant. On substituting the values of $\kappa_t$ from the experimental data at $x = 0.05$ m and $x = 0.30$ m, with we find that $\mu = 215$ m$^{-1} $.
 
\begin{figure}[h!]
				\centering
				\includegraphics[width=0.45\textwidth]{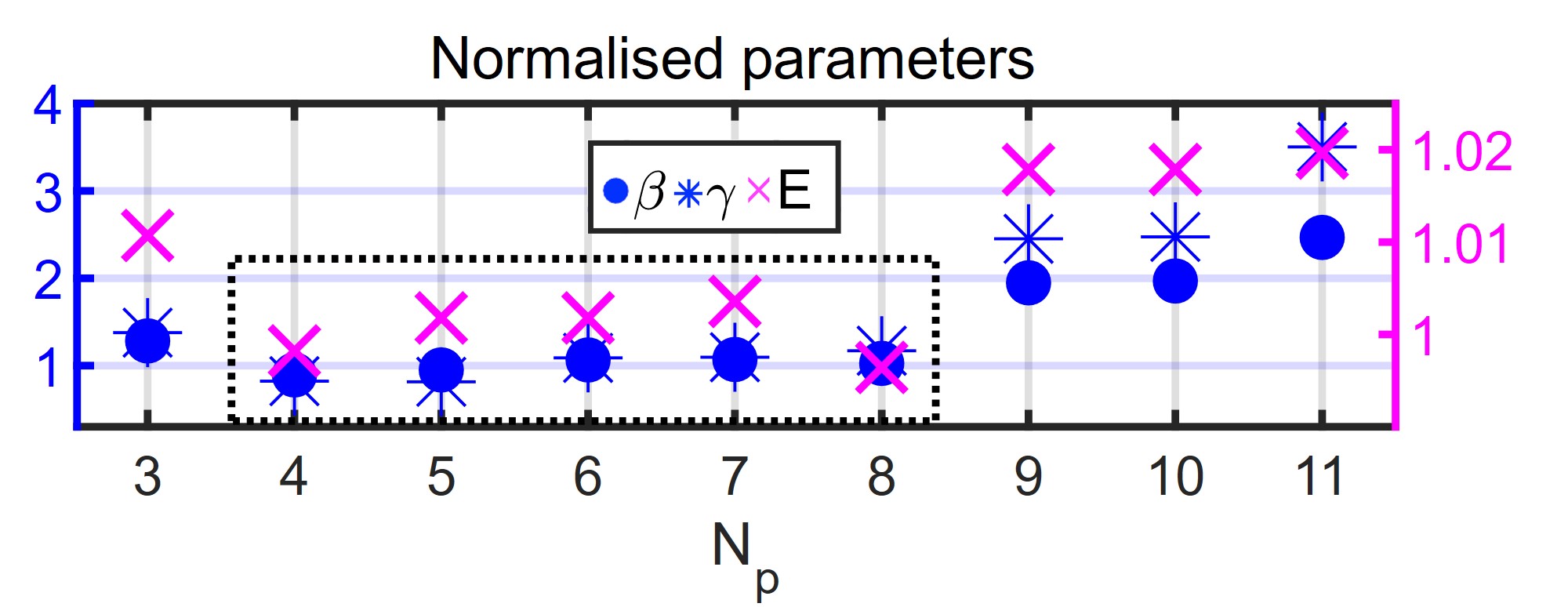}
				\caption{ The values of $E$ (right axis), $\beta$ and $\gamma$ (left axis)  on splitting the region $[0.30, 0.45]$ by $N_p$ points from induced tensile fracture. Values are normalised against  $E_G = 5.6$ GPa, $\beta_G = -38.5$ GPa and $\gamma_G = 1350$ GPa.}
				\label{fig:fit_blade}
			\end{figure}

		The same method as previous was used to fit $E$, $\beta$, $\gamma$, and then the parameters $a_1$ and $a_2$. The values used to construct the initial profile \eqref{eqn:IC} appropriate to this experiment were $\kappa = 0.022$, $\kappa_t = 0.0048$, $\eta_1 =  2.387 \times 10^{-5}$ s$^{-1}$ and $\tilde{L} = 2.1 \times 10^{-6}$ s$^{-1}$. The results for $E$, $\beta$, $\gamma$ are shown in Table \ref{tab:fit_blade} and Fig. \ref{fig:fit_blade}. The parameters reach a flat minimum around the points $N_p = 4,5,6,7,8$ as indicated by the dashed box in Fig. \ref{fig:fit_blade}, before increasing to relatively larger values. We take the average of those results as our parameters for this regime, giving $E_{G_2} = 5.6$ GPa, $\beta_{G_2} = -38.5$ GPa and $\gamma_{G_2} = 1350$ GPa. Then we determine $a_1 =   2\times 10^{-28}$ s$^5$ m$^{-1}$ and $a_2 = 4.9\times 10^{-15}$ s$^3$ m$^{-1}$. 
		
				\begin{table}
			\centering
				\begin{ruledtabular}
			\begin{tabular}{cccc}			
				\multicolumn{1}{l}{\textbf{$N_p$ \hspace{0.5cm}  }} & \multicolumn{1}{l}{\textbf{$E$ (GPa)\hspace{0.5cm} }} & \multicolumn{1}{l}{\textbf{$\beta$ (GPa)\hspace{0.5cm} }} & \multicolumn{1}{l}{\textbf{$\gamma$ (GPa)}} \\ \cline{1-4}
				3     \hspace{0.5cm}  & 5.66 \hspace{0.5cm}                                         &-49.4      \hspace{0.5cm}                                     &1860 \hspace{0.5cm}        \\    
				4        \hspace{0.5cm}                                        &   5.59 \hspace{0.5cm}                                      &-34.4      \hspace{0.5cm}                                     & 1110  \hspace{0.5cm}        \\   
				5       \hspace{0.5cm}                                                & 5.61   \hspace{0.5cm}                                       & -36.6   \hspace{0.5cm}                                      &1100  \hspace{0.5cm}        \\
				6   \hspace{0.5cm}                                              & 5.61    \hspace{0.5cm}                                     & -41.0    \hspace{0.5cm}                                     &1470   \hspace{0.5cm}       \\           
				7   \hspace{0.5cm}                                              & 5.62    \hspace{0.5cm}                                     & -41.2      \hspace{0.5cm}                                    &1480  \hspace{0.5cm}        \\  
				8   \hspace{0.5cm}                                               & 5.58    \hspace{0.5cm}                                     & -39.5       \hspace{0.5cm}                                  &1580   \hspace{0.5cm}        \\    
				9   \hspace{0.5cm}                                               & 5.70     \hspace{0.5cm}                                     & -75.0       \hspace{0.5cm}                                  &3310   \hspace{0.5cm}        \\   
				10   \hspace{0.5cm}                                               & 5.70     \hspace{0.5cm}                                     & -75.8        \hspace{0.5cm}                                  &3340  \hspace{0.5cm}        \\        
				11   \hspace{0.5cm}                                               & 5.71     \hspace{0.5cm}                                     & -95.0      \hspace{0.5cm}                                  &4730   \hspace{0.5cm}        \\ 		                             
			\end{tabular}
			\caption{The averaged values of $E$, $\beta$ and $\gamma$ found from fitting with different numbers of points $N_p$ using averaged 0.05 m and 0.10 m profiles from induced tensile fracture.}                                       
			\label{tab:fit_blade}
				\end{ruledtabular}
			\end{table}

		\begin{table}
			\centering
				\begin{ruledtabular}
			\begin{tabular}{ccc}			
				\multicolumn{1}{l}{\textbf{Parameter \hspace{0.5cm}  }} & \multicolumn{1}{l}{\textbf{Tensile \hspace{0.5cm} }} & \multicolumn{1}{l}{\textbf{Induced }}  \\ \cline{1-3}
				$E_{lin}$ (GPa)     \hspace{0.5cm} & 5.05 \hspace{0.5cm}                                        &5.46       \hspace{0.5cm}                                               \\    
				$E$ (GPa)     \hspace{0.5cm} & 5.17 \hspace{0.5cm}                                        &5.6       \hspace{0.5cm}                                               \\    
				$\beta$ (GPa)       \hspace{0.5cm}                                        &   -74.4  \hspace{0.5cm}                                      &-38.5     \hspace{0.5cm}                                             \\   
				$\gamma$ (GPa)      \hspace{0.5cm}    & 5940  \hspace{0.5cm}                                       & 1350  \hspace{0.5cm}                                              \\
				$\mu$ (m$^{-1}$)  \hspace{0.5cm}                                              & $-$  \hspace{0.5cm}                                     & 215      \hspace{0.5cm}                                             \\           
				$a_1$ (s$^5$m$^{-1}$ )  \hspace{0.5cm}                                              & $1\times 10^{-28}$     \hspace{0.5cm}                                     & $2\times 10^{-28}$     \hspace{0.5cm}                                          \\  
				$a_2$ (s$^3$m$^{-1}$) \hspace{0.5cm}                                               & $5.7\times 10^{-14}$     \hspace{0.5cm}                                     & $4.9\times 10^{-15}$          \hspace{0.5cm}                                         \\				$a_3$ (s$^3$m$^{-1}$) \hspace{0.5cm}                                               & $-$     \hspace{0.5cm}                                     & $-$        \hspace{0.5cm}                                         \    		                             
			\end{tabular}
			\caption{A summary of the parameter values obtained from the natural and induced  tensile fracture experiments (strain rates $\sim800$ s$^{-1}$ and $\sim1000$ s$^{-1}$ respectively) using the 
				theory-based fitting discussed in this section.}                                       
			\label{tab:parameters}
				\end{ruledtabular}
		\end{table}
		
		 					\begin{figure*}[t]
			\centering
			\begin{subfigure}[t!]{0.45\textwidth}
				\includegraphics[width=8.3cm]{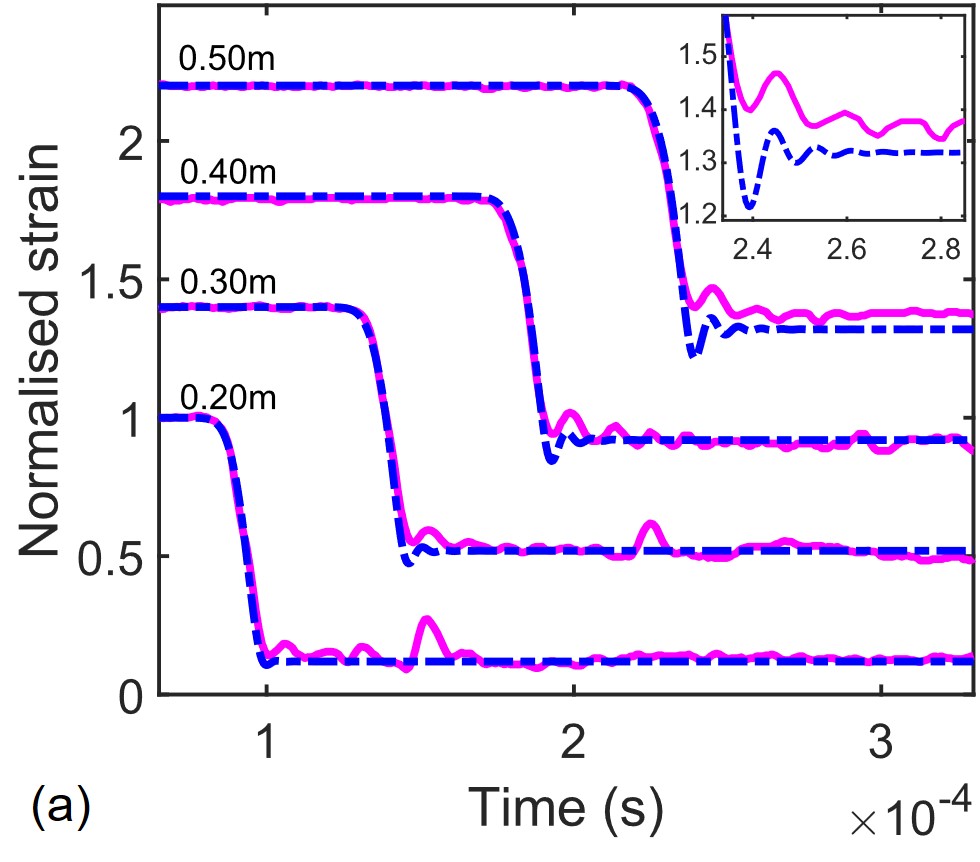}
				\label{fig:erf_G}
			\end{subfigure}	\ \
			~
			\begin{subfigure}[t!]{0.45\textwidth}
				\centering
				\includegraphics[width=8.3cm]{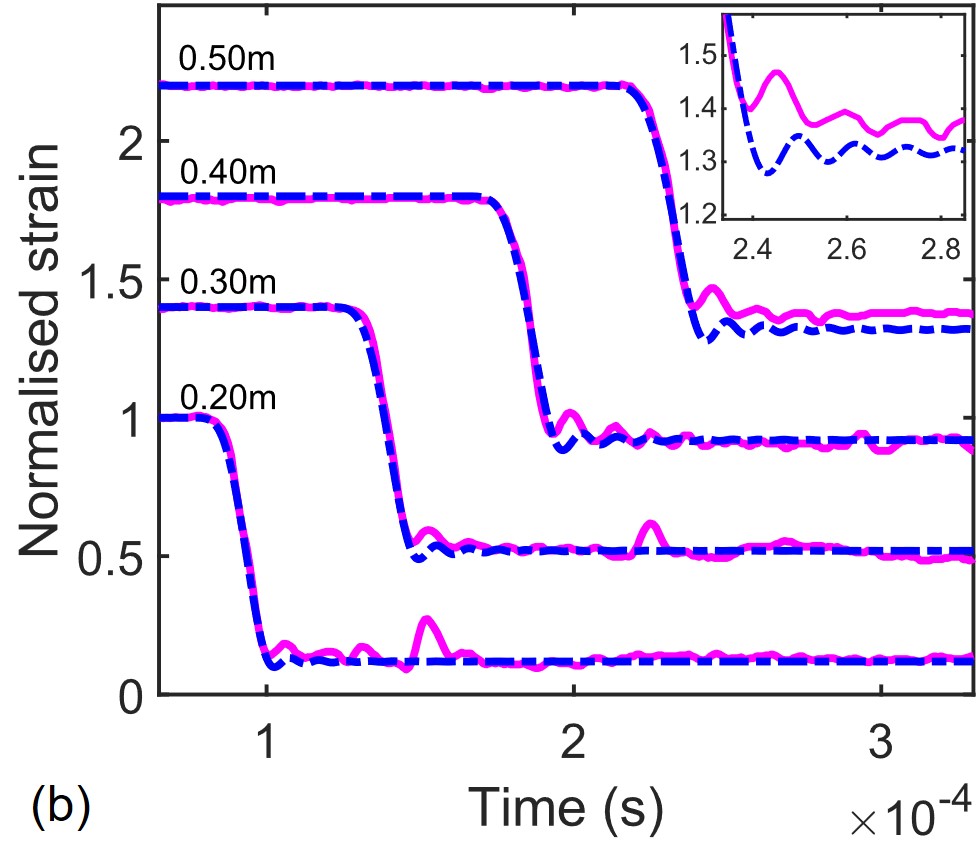}
				\label{fig:erf_L}
			\end{subfigure}
			\caption{(a) The (a) Gardner \eqref{Gardner} (blue, dash-dot), (b) eKdV (\eqref{5th_V_G} with $\mu = 0$) (blue, dash-dot) solutions with the error function initial profile given by \eqref{eqn:IC}, along with the experimental profiles (pink) between 0.20 - 0.50 m from the fracture site for natural tensile fracture. Profiles are normalised against the pre-strain and a vertical offset of 0.4 is used to separate each successive profile.
				\label{fig:models}
			}
		\end{figure*} 
		
		We note that the inclusion of viscoelastic effects introduces attenuation and changes the dispersion of the system. Therefore the correction provided by the  fifth derivative term should be viewed as accounting, at least partially, for both sources of dispersive corrections: geometric and viscoelastic, while attenuation is accounted for directly by the inclusion of the leading-order viscoelastic term into the model. 
		
		
		
		Again, our assumptions can now be validated as we find that $\epsilon = 0.151$, $\bar \beta = -1$, $\bar{\gamma} = 0.257$, $\bar{\delta}^2 = 0.005$, $\bar{a_1} = -0.0001$ and $\bar{a_2} = -0.01$ ($e_0 \sim 0.022, L_0 \sim 0.03$ m).
		The coefficient of the viscoelastic term in non-dimensional variables \eqref{ndv} is
		\begin{equation*}
			\bar{\mu} = \frac{\mu L_0 E}{2c_0e_0\lvert \beta \rvert},
		\end{equation*}
		which takes the value $\bar{\mu} = 0.009$ with the parameters above. 
		
The fitted value of $E$ obtained from the linear Gardner model was found to be $E_{{lin}_2}$ = 5.46 GPa. This value is again close to $5.42$ GPa, which is the value of Young's modulus 
obtained by linearising the Gardner equation \eqref{Gardner} about the average pre-strain at 0.10 m from the induced tensile fracture experiment. 
 The pre-strain and strain rate encountered in induced fracture experiments are greater that that in tensile fracture experiments and result in a more nonlinear regime.
                                                                                                 
		A summary of all parameters obtained is shown in Table \ref{tab:parameters}.		
		The parameters obtained from our two types of experiments are similar but not identical. 
	 Indeed, the strain rate encountered in the two experiments are different, therefore the parameter values obtained from them can also be expected to be different as the properties of PMMA are well known to be strain rate dependent. This is because, while  the leading order viscoelastic term, with a single time constant, can account
	for the experimentally observed rising level of post-strain and damping of the oscillations, full treatment of the viscoelastic behaviour of PMMA would require a model that includes a broad spectrum of time constants. In the approach to the description of viscoelasticity adopted in this paper, 
	the elastic parameters are strain rate dependent.

		 On comparison of $E_G$ and $E_{G_2}$ obtained from experiments with strain rates of $\sim 800 $ s$^{-1}$ and $\sim 2000$ s$^{-1}$ respectively, we note that $E_{G_2}$ is larger that $E_G$ which agrees with other experiments where Young's modulus of PMMA has been found to increase with strain rate \cite{WMX, Moy_2011}.


\section{Discussion}
 Results of the Gardner model \eqref{Gardner} and eKdV equation (\eqref{5th_V_G} with $\mu = 0$) 
	are shown individually in detail in Fig. \ref{fig:models} alongside corresponding experimental profiles from natural tensile fracture.  
	The 
	model equations were solved numerically with a pseudo-spectral method using 4000 points. The fourth-order Runge-Kutta method was used with the spatial step of $\displaystyle{1\times 10^{-4}}$~m
		and a rising slope was added sufficiently far away to the left 
		by the introduction of another error function for periodicity.

	Both models capture, with differing accuracy, the main features of the developing bore under the conditions of our experiment 
	with the parameters obtained from the fitting. In particular from Fig. \ref{fig:models}(a) the leading oscillation is reasonably well described by the Gardner model.	However it is clear that the 5th order dispersive term is required to capture further oscillations in the tail of the bore (Fig. \ref{fig:models}(b)).

		In both cases, the nonlinear Gardner models provide a better comparison to experiments than the linear equations obtained on linearisation near the pre-strain, especially at the later distances (0.40 - 0.50 m) where the slope of the linear solutions become too steep. 
The oscillatory part of the bore is similar between the linear and nonlinear Gardner models. In particular, the leading oscillation is reasonably well described by the linearised Gardner model which we discuss later. 
According to \cite{W_2019}, the conditions of the natural tensile fracture experiment in which the strains did not exceed much the value of 0.01 and strain rate $\sim 800$~s$^{-1}$, constitutes a weakly-nonlinear regime.
We also note that the stress at fracture was equal to $\sim 33$~MPa, which at our loading strain rate of $3 \times 10^{-3}$~s$^{-1}$ is also a borderline regime between linear and nonlinear (see Fig. 3 in \cite{WMX}). 
	
	In all models, the time at which the strain starts to decrease corresponds remarkably well to experimental results at each distance which indicates that the speed of the front of the release wave is captured well. However the nonlinear model continues to closely follow the slope of the release wave. The linear models separate towards the bottom.



		
				
						\begin{figure*}[t]
	\centering
	\begin{subfigure}[t!]{0.45\textwidth}
		\includegraphics[width=8.3cm]{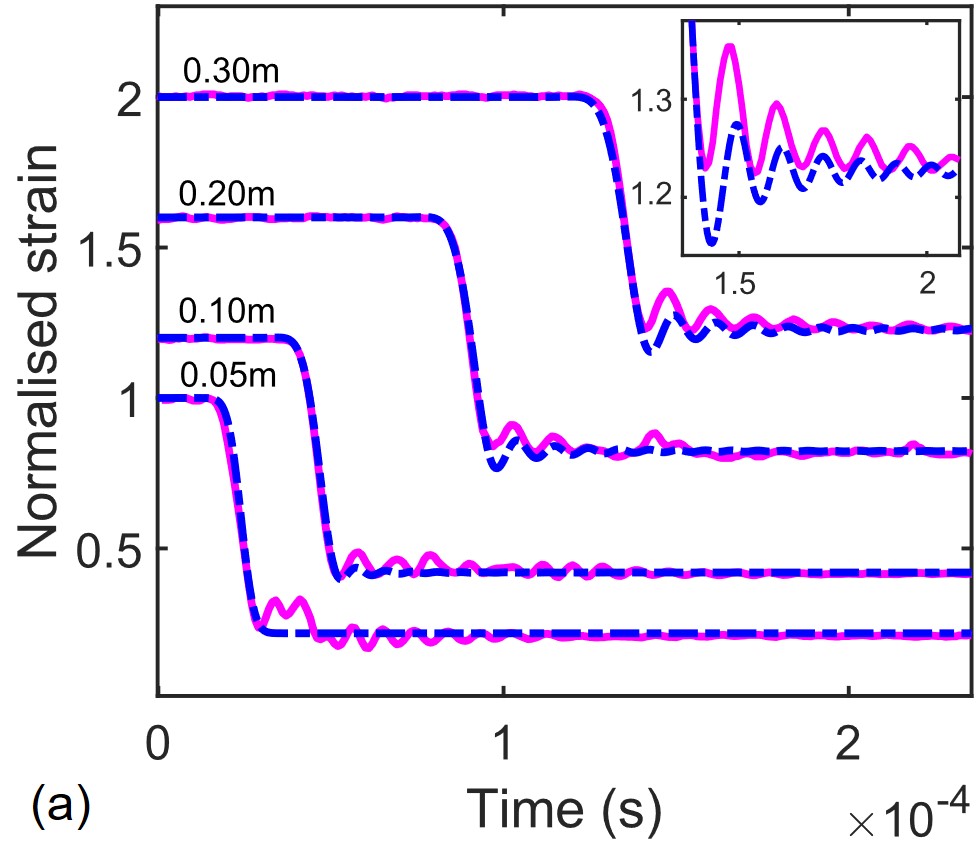}
		\label{fig:blade_erf}
	\end{subfigure}	\ \
	~
	\begin{subfigure}[t!]{0.45\textwidth}
		\centering
		\includegraphics[width=8.3cm]{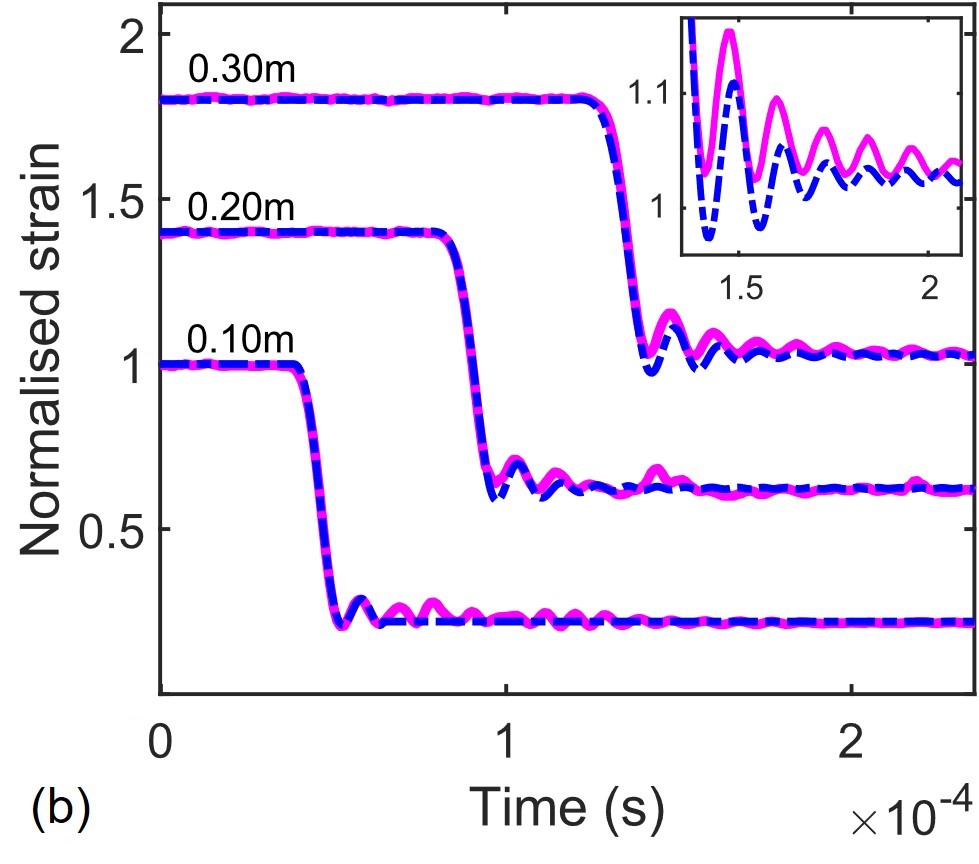}
		\label{fig:blade_spline}
	\end{subfigure}
	\caption{The veKdV model \eqref{5th_V_G} (a) with the error function initial profile at 0.05 m (blue, dash-dot)  and (b) with the spline initial profile at 0.10 m with experimental profiles (pink, solid) between 0.05 - 0.30 m from the induced tensile fracture site. Profiles are normalised against the pre-strain and a vertical offset of 0.4 is used to represent an increase of 0.10 m.}
	\label{fig:erf_spline}
\end{figure*}

We now turn attention to the induced tensile fracture experiment. 
	The initial profile given in the form of \eqref{eqn:IC} disregards any oscillations that are present at 0.05 m from the fracture site since they are masked by the shear waves. However, by 0.10 m there is sufficient separation between the longitudinal wave and the leading shear wave for the first oscillation to be identified. So to improve the initial profile, we used a spline approximation of the period from the top of the release wave to the end of the first oscillation of the 0.10 m induced tensile fracture experimental profile and used it as the initial profile. The profile was then scaled so that the pre-strain was equal to the average pre-strain of the experiments, 0.022, to best represent the entire set of results. An error function was again fitted sufficiently far away from the spline for periodicity, and the problem was solved numerically with a pseudo-spectral method as before.

 A comparison of the two initial profiles is given in Fig. \ref{fig:erf_spline} which shows the solutions of the veKdV model \eqref{5th_V_G} with parameters corresponding to the induced tensile fracture experiment (see Table \ref{tab:parameters}) with error function initial profile (Fig. \ref{fig:erf_spline}(a)) and a spline initial profile (Fig. \ref{fig:erf_spline}(b)), both  along with corresponding experimental profiles.
   
 Whilst the solution of the veKdV model with the value of $E_{G_2}$ fitted using our approximations did capture the bore quite well, a slight increase of $E_{G_2}$ from 5.6 GPa to 5.75 GPa gave a better overall agreement to the experimental data. We note that each measurement corresponds to an individual test with unique initial profiles, and we 
 find suitable parameters that best represent the entire set of results.

					\begin{figure*}[t]
	\centering
	\begin{subfigure}[t!]{0.45\textwidth}
		\includegraphics[width=8.3cm]{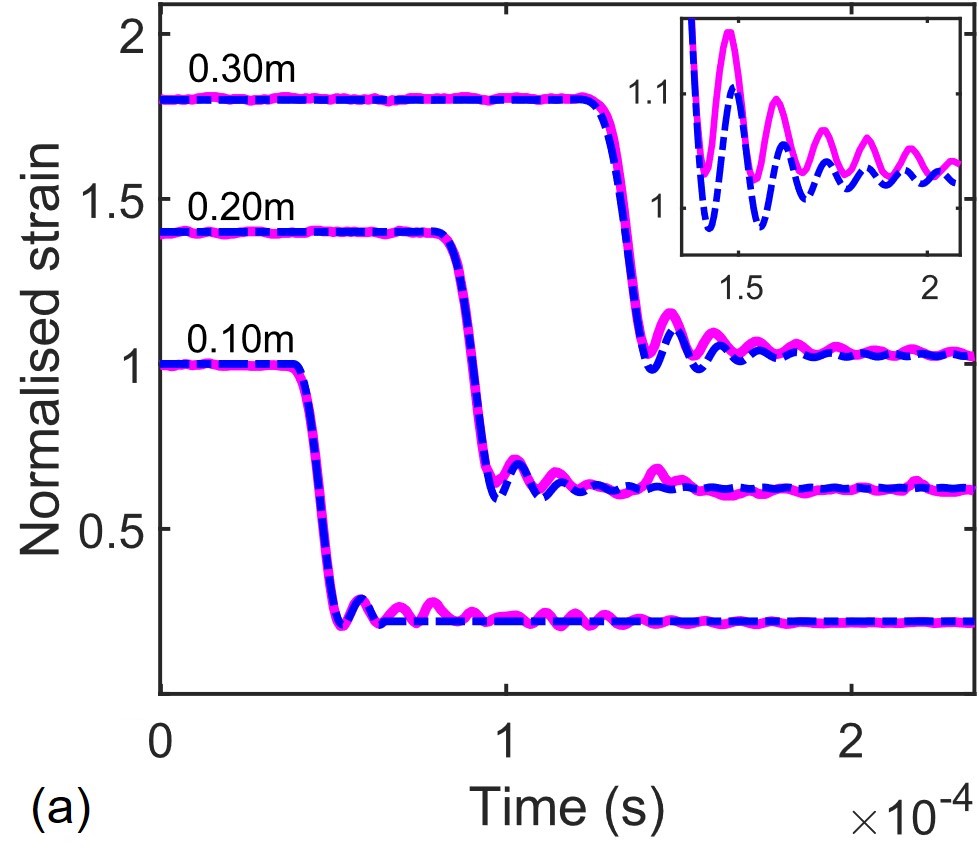}
		\label{fig:blade_G_spline}
	\end{subfigure}	\ \
	~
	\begin{subfigure}[t!]{0.45\textwidth}
		\centering
		\includegraphics[width=8.3cm]{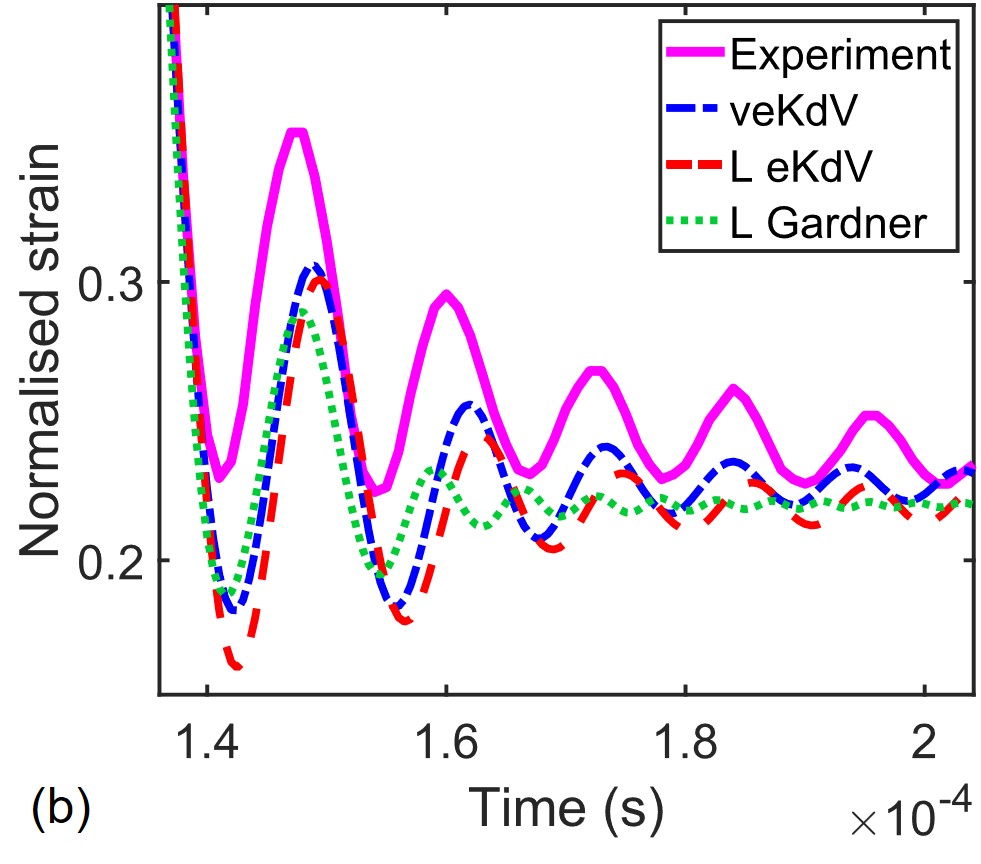}
		\label{fig:blade_G_spline_fit}
	\end{subfigure}
	\caption{The (a) veKdV model \eqref{5th_V_G} with spline initial profile at 0.10 m with $a_2$ fitted such that the amplitude of the first oscillation of the spline solution matches experiments at the 0.30 distance (blue, dash-dot) (b) a comparison of the linearised Gardner (green, dot), linearised eKdV, (red, dash) and veKdV (blue, dash-dot) models at 0.30 m from the induced tensile fracture site, all with the spline initial profile (0.10 m). Experimental profiles (pink, solid) between 0.10 - 0.30 m from the induced fracture site are also shown. Profiles are normalised against the pre-strain and a vertical offset of 0.4 is used to represent an increase of 0.10 m.}
	\label{fig:spline}
\end{figure*}

The slope of the release wave is captured well with both initial profiles. Oscillations develop from the initial profiles that gently increase in duration and amplitude with propagation distance which closely agrees with the oscillations observed in experiments at times before the arrival of the leading shear wave. The spline does give a better comparison at the later distance, and oscillations are very well described at all distances. 

The duration of the leading oscillation that emerges from the error function initial profile is the same as that in the solution from the spline at 0.30 m. The duration of the oscillations in the tail of the bore are slightly better captured with the spline initial profile (see inserts in Fig. \ref{fig:erf_spline}). 
 
 In Fig. \ref{fig:spline}(a) we show the solution of the veKdV model \eqref{5th_V_G} with the spline initial profile with the parameter $a_2 = 9 \times 10^{-15}$  s$^3$ m$^{-1}$ which was fitted so that the amplitude of the spline solution matched the experimental amplitude at 0.30 m. Improvements can be seen as the level of the first minimum is closer to the experimental first minimum with the spline. Also, the level of the first minimum in relation to the second minimum is better captured now.  
   
 A comparison of the linearised Gardner, linearised eKdV and veKdV models with the spline initial profile are shown in Fig. \ref{fig:spline}(b) along with the experimental profile at 0.30 m from the fracture site. 
 
 The effect of the 5th order dispersion term is clear on comparison of the linearised Gardner solution to the linearised eKdV solution. The tail of the bore is significantly extended with amplitude and duration of oscillations comparable to what is observed in the experiment. 
 
 The improvement due to the viscoelastic term can be seen in this comparison as the level on which the oscillations develop is in agreement with the experiment. The spline provides the best solution as it gives the closest match to the experiments.
 
 Although the amplitude of the leading oscillation of the linear solution is too small compared with experiment, the duration of the leading oscillation is captured rather well and is in good agreement with both experiments. Therefore, the linear bore solution discussed in the next section is relevant to our experiments, as well as being directly applicable to the modelling of undular bores in nearly linearly elastic materials such as steel.

\subsection{Linear Bore}	
	With the error function initial profile \eqref{eqn:IC}, this linear problem has the form
	\begin{equation}
		\begin{cases}
			\displaystyle{e_x + \frac{1}{c_0}e_t - \frac{\delta^2}{2c_0^3} e_{ttt} = 0,} \\
			\displaystyle{e(x_0,t) =  \kappa_t + \frac{\kappa-\kappa_t}{2}\left [ 1 - \text{erf} \lb \frac{t-\eta_1}{2 \tilde{L}} \rb \right ]}.
			\label{linear}
		\end{cases}
	\end{equation} 
	The transformation
	\begin{equation*}
		\tilde X = \frac{6 \delta^2}{2c_0^3}x, \hspace{0.6cm} \tilde T = t - \frac{x}{c_0}, \hspace{0.6cm} \bar{e} = \frac{\kappa - e}{\kappa - \kappa_t},
	\end{equation*}
	maps the problem \eqref{linear}  to that given in \cite{berry_2019} 
	(up to the change in notations). 
	We used the analytical solution 
	given in \cite{berry_2019} to obtain the solution to \eqref{linear} as
	\begin{align}
		&e (x,t) = \kappa_t + (\kappa - \kappa_t)\Big[ 1-\text{exp}\left(\frac{2L_1^6}{27x^2}\right) \times \nonumber \\ &\int_{b(x, t)}^\infty\text{exp} \left(\frac{sL_1^2}{(3x)^{2/3}}\right)\text{Ai}\left( s + \frac{L_1^4}{(3x)^{4/3}} \right) \text{ds} \Big], 
		\label{eqn:my_erf_sol}
	\end{align}
	where  $\displaystyle{b = \lb \frac{2}{3 \delta^2 x} \rb ^{1/3} \left [ x - c_0 \lb t-\eta_1 \rb \right ] }$, $\displaystyle{L_1 = \tilde{L}\left( \frac{2c_0^3}{\delta^2} \right)^{1/3}}$ and Ai is the Airy function.
	
	We note that the function $\tilde e (x, t) = e (x, t) - \kappa$ with $\kappa_t = 0$ is an analytical description of a compressive bore in impact experiments, within the scope of linear elasticity. 

	In the limit $x \to \infty$, the solution (\ref{eqn:my_erf_sol}) tends to the integral Airy solution (solution obtained when the initial condition is a step-function \cite{WT}) 	
	\begin{equation}
		e (x,t) = \kappa_t + (\kappa - \kappa_t)\Big[ 1- \int_{b(x, t)}^\infty\text{Ai}\left( s \right) \text{ds} \Big],
		\label{eqn:int_airy}
	\end{equation}	
	regardless of the gradient of the initial condition, i.e. the linear solution forgets its initial slope as it propagates \cite{berry_2019}. The solution \eqref{eqn:int_airy} is also obtained from \eqref{eqn:my_erf_sol} by taking the limit $\tilde{L} \to 0$, where by the initial condition tends towards a step between the levels of $\kappa_t$ and $\kappa$.
	
	The amplitude of the oscillatory part of the solution \eqref{eqn:my_erf_sol}, characterised by the vertical distance between the first minimum and first maximum, will grow and approach the limiting value as $x \to \infty$. Using the result for this limiting value in \cite{berry_2019}, we find that $a$
	approaches the limiting value of
	\begin{equation*}
		a \approx 0.466(\kappa - \kappa_t).
		\label{eqn:amp}
	\end{equation*} 
	Here, $\kappa - \kappa_t$ defines the difference between the values of the pre- and post- strain.

	As this value  is only achieved in the limit $x \to \infty$, it is not possible to give a distance from $x_0$ at which this value is reached, however the distance from $x_0$ at which the value of $a$ reaches a particular percentage of the maximum can be calculated. 
	We remove dependence on the slope parameter $\tilde{L}$, and the factor of $1/6$ in front of the dispersive term 
	by a further scaling
		$ \displaystyle \hat{T} = \frac{\tilde T}{\tilde{L}}, \ \hat{X} = \frac{\tilde X}{6 \tilde{L}^3}. $
	The solution to this reduced problem is given by 
	\begin{align*}
		&\bar{e}(\hat{X}, \hat{T}) = 1 - \text{exp} \lb \frac{2}{27\hat{X}^2} \rb \times \nonumber \\
		& \int_{b \lb \hat{X}, \hat{T} \rb}^{\infty} \text{exp} \lb \frac{s}{ \lb 3\hat{X} \rb ^{2/3}} \rb \text{Ai} \lb s + \frac{1}{\lb 3\hat{X} \rb ^{4/3}} \rb   ds,
		\label{eqn:sol_fixed}
	\end{align*}
	where $\displaystyle{b(\hat{X}, \hat{T}) = - \frac{\hat{T}}{ (3\hat{X})^{1/3}}  }$. 
	In the spirit of \cite{CS}, by numerically finding the distance at which the value of $a$
	reaches $50\%$ and $90\%$ of its limiting value, we find that the distance at which \eqref{eqn:my_erf_sol} reaches these thresholds as
	\begin{equation*}
		x_{50} \approx 3.5 \frac{2c_0^3 \tilde{L}^3}{\delta^2} ,
		\label{eqn:50perc}
	\end{equation*}
	and
	\begin{equation*}
		x_{90} \approx 56 \frac{2c_0^3 \tilde{L}^3}{\delta^2}  ,
		\label{eqn:90perc}
	\end{equation*}
	respectively.
	Note that here, smaller values of $\tilde L$ give steeper slopes, thus for steeper slopes, the 
	amplitude thresholds are reached sooner. With an increasing leading amplitude naturally comes more developed oscillations trailing behind. This helps to explain the reduced amplitude of oscillations in the 0.45 m profile of Fig. \ref{fig:exp_layer_comp} when compared to the 0.40 m profile for example, as the strain rates here were $\sim 700$ s$^{-1}$ and $\sim 800$ s$^{-1}$ respectively. 
		
		\begin{figure}
			\centering
			\includegraphics[width=8cm]{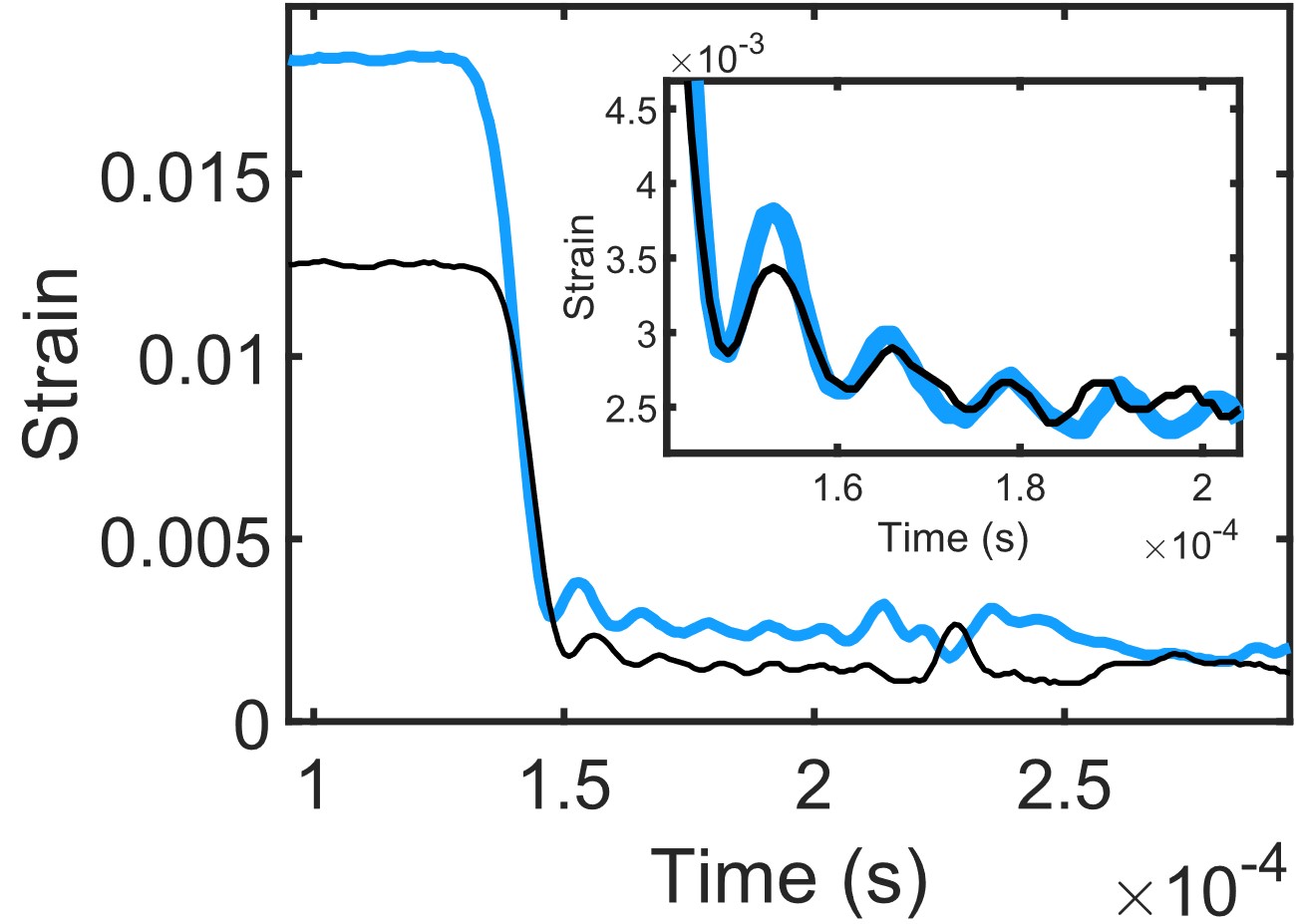}
			\caption{Two experimental strain profiles with different values of pre-strain recorded 0.30 m from the natural tensile fracture site. In the insert, the black (thin) curve has been translated in order to provide an easier comparison of oscillations. }
			\label{fig:300mm}
		\end{figure}


	The duration of the first oscillation of the bore in seconds, defined as the time between the first two minima of the solution \eqref{eqn:my_erf_sol}, can be given by finding its stationary points. Indeed, differentiating \eqref{eqn:my_erf_sol} with respect to $t$ gives 
	\begin{align}
		e_t(x,t) &= c_0 (\kappa - \kappa_t) \text{exp} \lb \frac{2 L_1^6}{27 x^2}\rb \text{exp} \lb \frac{b L_1^2}{ \lb 3x \rb ^{2/3}} \rb \times \nonumber \\ 
		&  \text{Ai} \left( b + \frac{L_1^4}{(3x)^{4/3} }  \right) \lb \frac{2}{3 \delta^2 x} \rb ^{1/3} .
		\label{eqn:e_t}
	\end{align}
	On solving $e_t (x, t) = 0$, it is clear that this is only satisfied at the zeros of the Airy function. In particular, we are interested in the 1st and 3rd zeros, which correspond to the first and second minima, to calculate the duration. These are found when the argument of the Airy function is equal to $-2.33811$ and $-5.52056$ respectively \cite{Airy}. From this we find the duration as the difference between the values of the  first and third stationary points to be gently increasing as	
	\begin{align*}
		\hat t (x) 
		& \approx \frac{3.643}{c_0} \lb \delta^2 x \rb ^{1/3}.
		\label{eqn:dur}
	\end{align*} 
	This is an extension of the result given in \cite{berry_2019} for the integral Airy solution. 
	Importantly, this result is not dependent on the height or gradient of the initial slope as there is no dependence on $\kappa$, $\kappa_t$ or $\tilde{L}$. Thus the expansion of the duration is the same for the integral Airy solution \eqref{eqn:int_airy} and the solution \eqref{eqn:my_erf_sol}. The independence of duration on strain rate, pre-strain and post-strain can be seen from the natural tensile fracture profiles presented in Fig. \ref{fig:300mm}. Indeed, the pre-strain, post-strain and strain rate are different between the two measurements, as well as the difference between pre-strain and post-strain, but the duration of the first oscillation is the same in both. 	
	As a final comment regarding Fig. \ref{fig:300mm}, one should note the clear structure of an undular bore in both experimental profiles. 
	
	

	While the gradient of the solution is given in full by \eqref{eqn:e_t}, it is instructive to find the gradient at a point on the front slope of the bore by evaluating it at $t = \eta_1 + x/c_0$, where $x$ is the distance from $x_0$. On substitution, we have 
	\begin{align*}
		g_s(x) & =  c_0 (\kappa - \kappa_t)\text{exp} \lb \frac{2 L_1^6}{27 x^2}\rb \times \nonumber \\
		& \text{Ai}  \lb \frac{ L_1^4}{(3x)^{4/3}} \rb   \lb \frac{2}{3 \delta^2 x} \rb  ^{1/3}.
	\end{align*}
	This function gradually decreases with propagation distance, thus the front slope of the bore gets less steep.  This formula is an extension of the counterpart of the result for the integral Airy solution in \cite{berry_2019}, which can be recovered by taking the limit $\tilde{L} \to 0$ as
	\begin{align*}
		g_s(x) & \approx  0.3101 c_0 (\kappa - \kappa_t) \lb \frac{1}{ \delta^2 x}  \rb ^{1/3}.
	\end{align*}
	


\section{Conclusion}
	Features of an undular bore with a gently expanding in amplitude and duration oscillatory structure connecting two levels of strain have been observed, for the first time, in two sets of laboratory experiments with a PMMA waveguide undergoing relaxation after both natural and induced tensile fracture.

A viscoelastic extended KdV equation was derived which is able to capture the main features of the observed structure at times up to the arrival of the leading shear wave.
 
		Simple formulae to describe features such as the amplitude and duration of the leading oscillation have been derived from the analytical solution to the linearised near pre-strain Gardner model which provide a qualitative agreement to what is observed in experiments. This observation invites theoretical studies within the scope of the extended viscoelastic models in order to investigate the full range of validity of this approximation under the conditions of our experiments. Useful reviews of the dissipative extensions of the KdV and Gardner-type models of undular bores can be found in \cite{Kamch,El_2017}.
		
The 5th order dispersive term captures additional oscillations in the tail of the bore up to times of the arrival of shear waves which are not modelled. Coupled models for longitudinal and shear waves should be used in this area. 
	
Some features, such as the rising level of post-strain with distance from fracture, 
are not described by the elastic models but have been captured by the introduction of the leading-order viscoelastic term.
	

	To fit the parameters more accurately, experimental measurements of the same wave at different distances is required. We reiterate that all measurements presented in this paper are from individual tests which will all, inevitably, have slightly different initial conditions. The parameters that we have obtained have proven to give good agreements between the models and experiments and are in line with other experiments that have been performed on PMMA at strain rates similar to what we have presented.

	Experiments with longer waveguides to enable measurements further away from fracture 
	are necessary to determine the long-time development of undular bores in PMMA, and the limitations of the models. The research opens new avenues for the study of undular bores in solids, including their interaction, reflection and scattering by defects, 
	with possible applications in non-destructive testing.

	From the studies of undular bores described by the KdV / Gardner-type equations, and their dissipative extensions  \cite{GP, Gbore, MS_06, SH,  El_2017, BS}, the behaviour of the bore strongly depends on the initial strain and slope of the wave and may result, for greater values of both, in a significant increase of the amplitude of oscillations.
Such waves could be present in the signals generated by earthquakes, fracking and other similar events in situations involving transverse fracture of an appropriately pre-strained waveguide.


	

	\section*{Acknowledgements} 
	We thank Professor Fabrice Pierron and Dr. Lloyd Fletcher for their experiments and use of the grid method to identify the shear wave. 
	K.R.K. thanks Professor Sir Michael Berry, Professor Noel Smyth and Dr. Dmitri Tseluiko for useful discussions. The authors acknowledge the funding from the School of Science and WSMEME at Loughborough University. C.G.H. is grateful to the WSMEME for a School Studentship. 

	\medskip
	
	 {\it Corresponding author:} K.Khusnutdinova@lboro.ac.uk 

\end{document}